\documentclass[%
 reprint,
 amsmath,amssymb,
 aps,
]{revtex4-2}

\usepackage{CJK}
\usepackage[utf8]{inputenc}
\usepackage{graphicx} % Include figure files
\usepackage{dcolumn}  % Align table columns on decimal point
\usepackage{bm}       % bold math
\usepackage[caption=false,font=footnotesize]{subfig}
\usepackage[dvipsnames]{xcolor}
\usepackage{url}
\usepackage[hidelinks]{hyperref}
\hypersetup{
  colorlinks = true, % Colours links instead of ugly boxes
  urlcolor   = blue, % Colour for external hyperlinks
  linkcolor  = blue, % Colour of internal links
  citecolor  = green, % Colour of citations
    pdfinfo  = {
      Title={Time-invariant degree growth in preferential attachment network models},
      Author={Jun Sun, Matúš Medo, Steffen Staab},
      Subject={Physics, Computer Science}
    }
}

\newcommand{\eg}{\emph{e.g.}}
\newcommand{\ie}{\emph{i.e.}}
\newcommand{\eref}[1]{Eq.~(\ref{#1})}
\newcommand{\rme}{\text{e}}
\newcommand{\der}{\text{d}}

\graphicspath{{./}}

\begin{document}
\begin{CJK*}{UTF8}{gbsn}

\title{Time-invariant degree growth in preferential attachment network models}

\author{Jun Sun (孙骏)}
\email{junsun@uni-koblenz.de}
\affiliation{Institute for Web Science and Technologies, Universit\"{a}t Koblenz--Landau, 56070 Koblenz, Germany}

\author{Mat\'{u}\v{s} Medo}
\email{matus-medo@unifr.ch}
\affiliation{Institute of Fundamental and Frontier Sciences, University of Electronic Science and Technology of China, Chengdu 610054, PR China}
\affiliation{Department of Radiation Oncology, Inselspital, Bern University Hospital and University of Bern, 3010 Bern, Switzerland}
\affiliation{Department of Physics, University of Fribourg, 1700 Fribourg, Switzerland}

\author{Steffen Staab}
\email{s.r.staab@soton.ac.uk}
\affiliation{Institute for Parallel and Distributed Systems, Universit\"{a}t Stuttgart, 70569 Stuttgart, Germany}
\affiliation{Web and Internet Science Research Group, University of Southampton, SO17 1BJ, UK}

\date[Submitted: ]{October 10, 2019; Revised: January 10, 2020}

\begin{abstract}
Preferential attachment drives the evolution of many complex networks. Its analytical studies mostly consider the simplest case of a network that grows uniformly in time despite the accelerating growth of many real networks. Motivated by the observation that the average degree growth of nodes is time-invariant in empirical network data, we study the degree dynamics in the relevant class of network models where preferential attachment is combined with heterogeneous node fitness and aging.
We propose a novel analytical framework based on the time-invariance of the studied systems and show that it is self-consistent only for two special network growth forms: the uniform and exponential network growth. Conversely, the breaking of such time-invariance explains the winner-takes-all effect in some model settings, revealing the connection between the Bose-Einstein condensation in the Bianconi-Barab\'{a}si model and similar gelation in superlinear preferential attachment. Aging is necessary to reproduce realistic node degree growth curves and can prevent the winner-takes-all effect under weak conditions. Our results are verified by extensive numerical simulations.

\end{abstract}
\maketitle
\end{CJK*}

\section{\label{sec_intro}Introduction}
The original work on the preferential attachment network growth mechanism~\cite{barabasi1999emergence} has importantly contributed to the formation of the interdisciplinary field of network science~\cite{barabasi2016network, newman2018networks}. Since then, preferential attachment-based network models have been used to model the evolution of a broad range of networks, such as the World Wide Web~\cite{barabasi1999emergence, adamic2000power}, citation networks~\cite{medo2011temporal, jeong2003measuring}, and social networks~\cite{newman2001clustering, capocci2006preferential}. The most important generalizations of the original preferential attachment model are the inclusion of the node-specific fitness parameter~\cite{bianconi2001competition} and aging that suppresses the attractiveness of old nodes to new links~\cite{dorogovtsev2000structure}.
The basic preferential attachment mechanism, also known as the rich-get-richer or the Matthews effect, dictates that nodes attract new links at a rate that is proportional to the degree that they already have. This microscopic mechanism induces a positive feedback loop that results in a network degree distribution that is power-law (scale-free) under some model settings~\cite{albert2002statistical}. Similar broad degree distributions, though seldom of an ideal power-law shape, are found in many real-world networks~\cite{newman2001clustering, clauset2009power}.

We build our work on the observation that in many real-world networks, using citation networks as an example here, the degree growth is time-invariant: the average degree of nodes of different age has the same functional dependency on node age regardless of when the nodes have entered the network. This seemingly minor observation is actually not trivial. First of all, preferential attachment models without aging are known to have a strong first-mover advantage: the first nodes accumulate many more links than the nodes that enter the network later~\cite{newman2009first}.
We show that an accelerated network growth, a feature that is common in real networks~\cite{dorogovtsev2001effect, parolo2015attention, gagen2005accelerating} yet usually overlooked by network modeling, is an important part of the interplay between preferential attachment and the macroscopic degree growth patterns. In particular, of different growth forms that can be considered, the exponential network growth is consistent with the time-invariant degree growth.

\begin{figure*}
  \centering
  \includegraphics[scale=0.5]{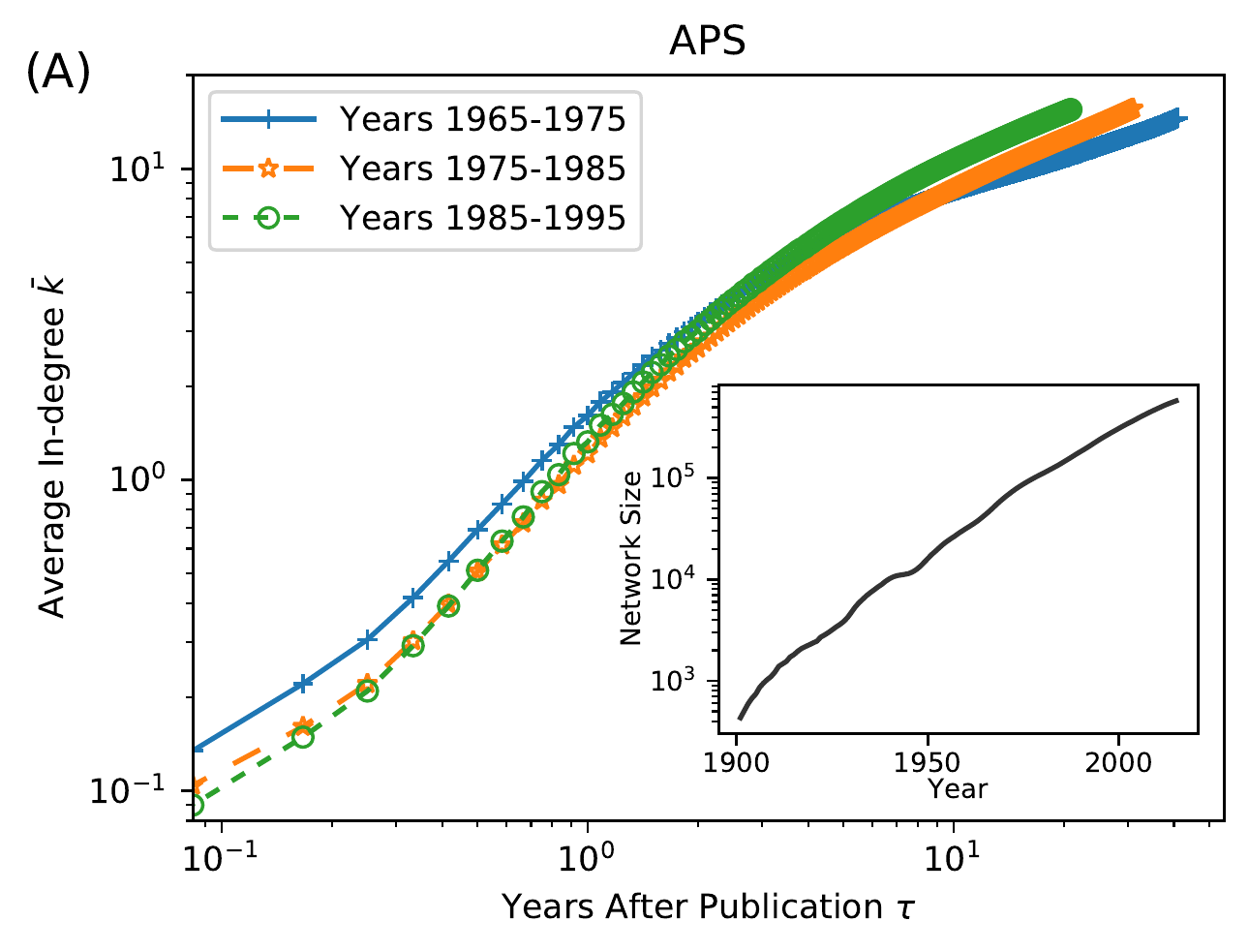}
  \quad
  \includegraphics[scale=0.5]{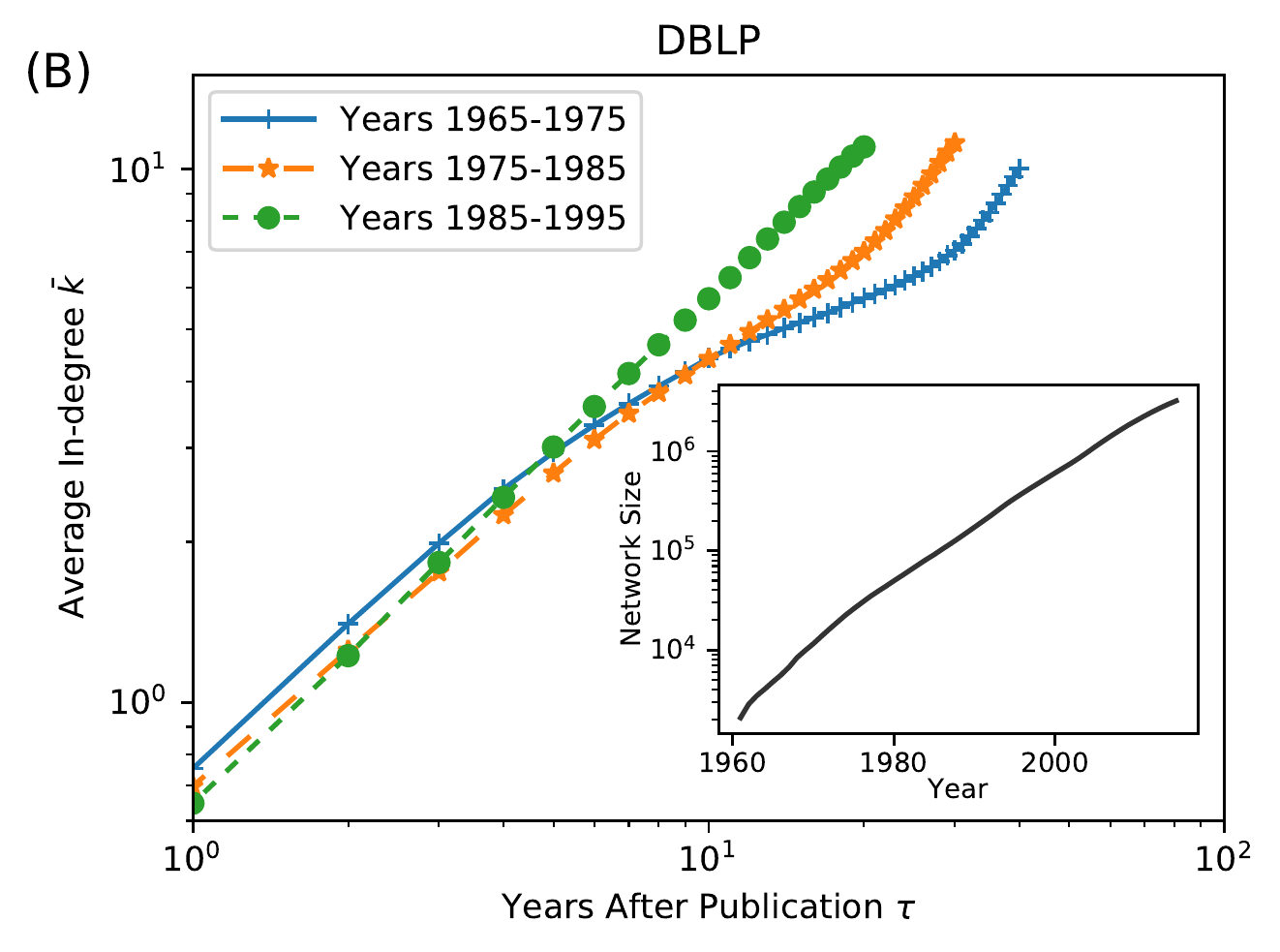}
  \caption{The average number of citations as a function of paper age for papers published in different time periods in (A) the APS data and (B) the DBLP data. Papers are grouped by their publication year. The insets show how the number of papers in each dataset grows with time. Note that the main plots use the log-log scale and the insets use the linear-log scale.}
  \label{fig_exp_growth}
\end{figure*}

To systematically explore the conditions under which a time-invariant degree growth arises, we introduce a novel mathematical formalism for preferential attachment-based models, where exponential and linear network growth emerge as the only possible solutions of an eigenvalue problem. The new formalism also reveals the connection between the Bose-Einstein condensation~\cite{bianconi2001bose} in the Bianconi-Barab\'{a}si model~\cite{bianconi2001competition} and a similar gelation phenomenon seen in the superlinear preferential attachment~\cite{krapivsky2000connectivity}. Aging~\cite{dorogovtsev2000structure, medo2011temporal} is necessary to recover realistic degree growth curves that are slower than exponential (\eg, power functions).

The paper is organized as follows. In Section~\ref{sec:empirical}, we present our empirical findings in real data and motivate our study. In Section~\ref{sec_model}, we introduce relevant network models, study their analytical properties, and introduce a mathematical framework for growing networks with time-invariant degree growth. In Section~\ref{sec_simulation}, we generate synthetic networks with different parameters to evaluate our analytical results. In Section~\ref{sec:conclusion}, we conclude with some discussions and point to potential future work.

\section{Empirical evidence}
\label{sec:empirical}
We begin by studying the growth patterns in real datasets. We use two citation networks in particular: the American Physical Society (APS) citation network (available from \url{https://journals.aps.org/datasets}),
and the computer science citation network extracted by Tang et al.~\cite{aminer} (available from \url{https://aminer.org/citation}), originally indexed by the DBLP computer science bibliography website~\cite{ley2002dblp}.
The APS dataset comprises 564,517 papers published in the APS journals from 1893 to 2015 and 6,715,562 citations among them. The DBLP dataset comprises 3,272,991 computer science papers published from 1936 to 2016 and 8,466,859 citations among them. The paper publication dates are available with the time resolution of one day and one year for the APS and the DBLP data, respectively. In the network representation, a citation between two papers corresponds to a directed link between two network nodes. The node out-degree is determined at the moment when the paper together with its list of references is published. By contrast, the node in-degree gradually grows from the initial zero value. In terms of growth, we thus focus here on node in-degree.

\begin{figure}
\includegraphics[scale=0.5]{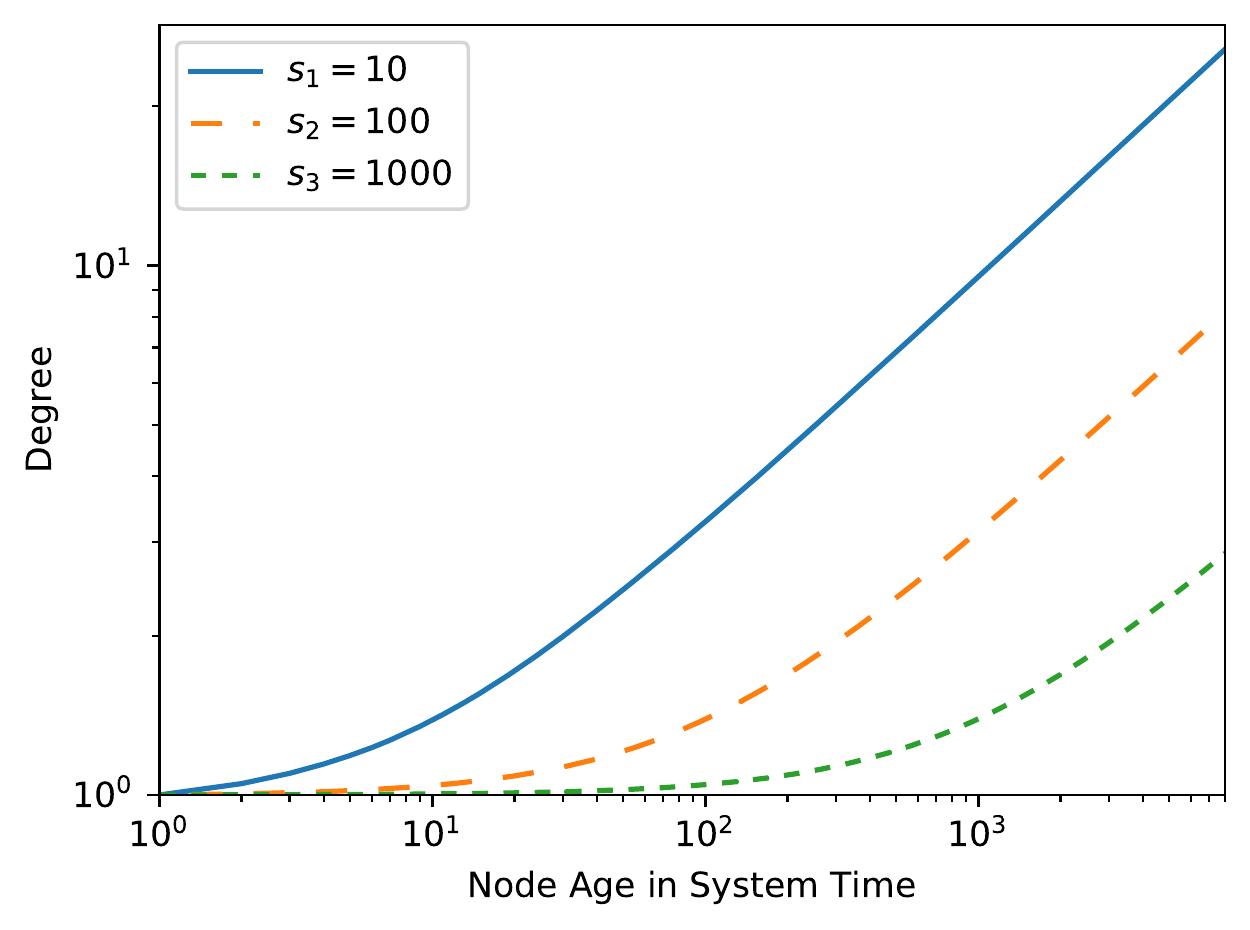}
\caption{The expected degree growth curves of three nodes which join the network at different times (with the same fitness value) in the Bianconi-Barab\'{a}si model.}
\label{fig_degree_growth_bianconi}
\end{figure}

In Fig.~\ref{fig_exp_growth}, we group the nodes by their publication date and plot the average in-degree as a function of the node age separately for nodes originating from different periods.
The average paper out-degree is now much higher than it was 50 or more years ago.
To limit the impact of this effect, we focus on the time period 1965--1995 during which the average out-degree of papers changed little (see Appendix).
Albeit the individual curves correspond to papers whose publication dates differ by up to 30 years, their shape is strikingly similar.
For the APS, we see various curves collapsing onto each other. This indicates that the manner in which the papers' average number of citations grow with paper age is time-invariant. While the curves' shapes are more complex for the DBLP data, they are still time-invariant for paper age less than approximately 20 years. 
In particular, old nodes do not have an advantage over the new ones, compared with Fig.~\ref{fig_degree_growth_bianconi} in which the early-mover advantage is forceful and, in turn, the growth of node degree is strongly determined by the time in which a node appears.
These results show that the in-degree growth function $k$ is time-invariant---it can be written as a function of the node age $\tau$ regardless of the node's appearance time.

Panels of Fig.~\ref{fig_exp_growth} further feature insets showing the evolution of the overall network size (measured by the number of nodes). They both point at an approximately exponential growth of the network size $s$ with time $t$, $s=\exp(\alpha t)$. Being non-linear to the \emph{physical time} $t$, the network size $s$ can be seen as the \emph{system time} driven by the arrival of new nodes. The non-linear relationship between $s$ and $t$ is however not considered in the original preferential attachment models. In this paper, we fill this gap and make a clear distinction between them.

\section{Time-invariant degree growth in network models}
\label{sec_model}
While the described real datasets are represented with directed networks, we focus here on undirected network models which attract more general interest than directed ones. The behavior of these two classes of models is often similar.

\subsection{The Bianconi-Barab\'{a}si model}
Before proceeding to more general considerations, we address here specifically the Bianconi-Barab\'{a}si model~\cite{bianconi2001competition} where preferential attachment is complemented with node fitness. The attractiveness of node $i$ to new links thus has the form $k_i\eta_i$ where $k_i$ and $\eta_i$ are the node degree and node fitness, respectively. Fitness~\cite{caldarelli2002scale} is an intrinsic property of a node. Nodes with higher fitness are more likely to attract links, thus their degrees tend to grow faster.
Node fitness is typically drawn from some probabilistic distribution $\rho(\eta)$ whose shape is an important constituent of the network model.
At the micro level, $\rho(\eta)$ allows nodes of the same age to grow at different rates. At the macro level, $\rho(\eta)$ affects the broadness of the resulting degree distribution~\cite{bianconi2001competition}.
For real data, the aim can be to determine node fitness values that best correspond to the observed data~\cite{medo2014statistical,wang127}.

The average degree growth in the Bianconi-Barab\'{a}si model has been shown~\cite{bianconi2001competition} to follow a power function
\begin{equation}
\label{eq_power_law_growth_s}
k_i(s, s_i, \eta_i) \sim (s / s_i)^{\beta(\eta_i)},
\end{equation}
where $s_i$ is the system time (network size) when node $i$ has appeared, $s$ is the current system time, and the exponent $\beta$ is a function of node fitness (for the basic model version, $\beta(\eta_i)\sim\eta_i$). As shown in Fig.~\ref{fig_degree_growth_bianconi}, such degree growth is clearly not invariant under the shift of the system time $s$. If, motivated by the exponential network growth size demonstrated in Fig.~\ref{fig_exp_growth}, we assume that $s=\rme^{\alpha t}$, \eref{eq_power_law_growth_s} is converted to
\begin{equation}
\label{eq_degree_exp}
k_i(t, t_i, \eta_i) \sim \rme^{\alpha \beta(\eta_i) (t-t_i)}
\end{equation}
where $t_i$ is the physical time when node $i$ has appeared and $t$ is the physical observation time. This form is indeed time-invariant as it depends on the node age $\tau_i:=t-t_i$ with no additional dependence on the node appearance time $t_i$.

We thus see that the Bianconi-Barab\'{a}si model produces a time-invariant degree growth if and only if the number of nodes grows exponentially with time. There is, however, still an important difference between the growth produced by \eref{eq_degree_exp} and the real data observations in Fig~\ref{fig_exp_growth}. While the former is of an exponential kind, the nearly linear curves in Fig.~\ref{fig_exp_growth} (log-log scale) suggest a power-law growth, much slower than the exponential growth. To resolve this disagreement, we proceed to more general preferential attachment models with fitness and aging~\cite{medo2011temporal} where the aging effect causes a slowdown of the degree growth.

\subsection{General preferential attachment with fitness and aging}
The general model that we aim to study has three main contributing factors:
\textit{node degree} as a classical amplifier that can be introduced by various mechanisms such as the reference-copying process~\cite{kleinberg1999web}, \textit{node fitness} as a reflection of intrinsic differences between the nodes, and \textit{aging} as a mechanism that reflects the natural preference for new and, at the same time, limits the strong bias towards old nodes.
The product of fitness and aging has also been referred to as ``relevance'' in past literature~\cite{medo2011temporal}.
The probability that node $i$ attracts a new link is usually assumed in the form $\Pi_i \sim k_i \eta_i R(\tau_i)$ where $\tau_i$ is the age of node $i$ (in physical time) and $R(\tau_i)$ is typically a decreasing function which represents the gradual loss of the node's ``relevance'' and contributes to an eventual saturation of the degree growth. It is convenient to set $R(0) = 1$ so that aging begins to influence the degree dynamics only later during each node's lifetime. Node fitness values are drawn from the distribution $\rho(\eta)$ which does not change with time. The number of nodes is assumed to grow exponentially with time, $s = \exp(\alpha t)$.

The continuum approximation for the degree evolution~\cite{dorogovtsev2000structure} replaces the stochastic evolution of each node's degree with the average rate of its increase, $\der k_i / \der s = m\Pi_i(s)$ where $m$ is the average number of new links created by a new node. Assuming that each new node creates \emph{one link} to an already existing node, we obtain
\begin{equation}
\frac{\der k_i}{\der s} = \frac{k_i(s) \eta _i R(\tau_i)}{Z(s)}
\end{equation}
where $Z(s):=\sum_j k_j(s)\eta_j R(\tau_j)$. It is convenient here to switch to the physical time $t$ where the rate equation has the form
\begin{equation}
\label{eq_degree_diff_func}
\frac{\der k_i}{\der t} = \alpha \rme^{\alpha t} \times \frac{k_i \eta _i R(\tau _i)}{Z(t)}
\end{equation}
When $t$ is large, the discrete sum in $Z(t)$ can be approximated with the double integral of the product $k_i \eta_i R(\tau_i)$ for all nodes, first over all possible node ages $\tau$, then over all possible fitness values $\eta$,
\begin{equation}
\label{eq_Z_simq}
Z(t) \approx \int\der\eta\, \rho(\eta)\eta \int_{0}^{t}\der\tau\, k(\tau, \eta) R(\tau) \times
\alpha \rme^{\alpha(t-\tau)}.
\end{equation}
Since the network size grows exponentially, there are more nodes with smaller ages in the network, thus the density term $\alpha \rme^{\alpha(t-\tau)}$ is used when integrating over node ages $\tau$.
Denoting
$$
\lim_{t\to\infty}\int\der\eta\, \rho(\eta)\eta\int_{0}^{t}\der\tau\, k(\tau, \eta) R(\tau)\rme^{-\alpha\tau}=\theta,
$$
we can write $Z(t) = \alpha\rme^{\alpha t}\theta$. \eref{eq_degree_diff_func} now simplifies to the form 
\begin{equation}
\label{deg_growth_exp_size}
\frac{\der k_i}{\der\tau_i} = \frac{k_i(\tau_i) \eta_i R(\tau_i)}{\theta}
\end{equation}
where we also used $\tau_i = t - t_i$ and replaced the derivative with respect to $t$ by the derivative with respect to $\tau$. The solution of this differential equation has the form
\begin{equation}
\label{eq_degree_growth}
k_i(\tau_i, \eta_i) = \exp\big[\eta_i r(\tau_i) / \theta\big]
\end{equation}
with $r(\tau):=\int_0^{\tau} R(t)\,\der t$. Since $r(0) = 0$, $k(0) = 1$ as expected. Different aging functions now lead to different forms of the degree growth $k$. 
In particular, the aging function $R(\tau) = (\tau + 1)^{-1}$ leads to a power-law degree growth $k(\tau, \eta) = \tau^{\eta/\theta}$ that can approximate the average degree growth in empirical data. In any case, \eref{eq_degree_growth} shows that this model together with the assumption of an exponentially growing network size produces time-invariant degree growth.

After the term $\alpha\rme^{\alpha t}$ introduced in \eref{eq_degree_diff_func} by the accelerating network growth being canceled with the same term in $Z(t)$, the implied differential equation for the degree growth, \eref{deg_growth_exp_size}, is the same as when the uniform network growth ($s = t$) is assumed~\cite{medo2011temporal}. In contrast to~\cite{medo2011temporal} where the normalization term $\sum_j k_j \eta_j R(\tau)$ converges only if $R(\tau)$ decays sufficiently fast (faster than $1/\tau$), we do not have a similar constraint here, as the exponential growth introduces the term $\rme^{-\alpha\tau}$ in $\theta$; this ensures convergence even when $R(\tau)$ decays no faster than $1/\tau$, for instance as in~\cite{sun2018decay}.

\subsection{Time-invariance of the degree growth as a required property}
We have shown that the model introduced in \cite{medo2011temporal} produces time-invariant degree growth when the network size grows uniformly or exponentially. We now proceed by showing that the uniform and exponential network growths are in fact the only two cases that are consistent with the time-invariant degree growth. To this end, we introduce a novel mathematical formalism for growing networks with the time-invariant degree growth as a fundamental assumption, but without an assumption on the network growth form in the first place.

To achieve a time-invariant degree growth for node attractiveness $\Pi_i \sim k_i \eta_i R(\tau_i)$, the differential equation of the degree growth function must take the form
\begin{equation}
\label{eq_degree_pde}
\frac{\der k_i}{\der\tau_i} = c k_i \eta_i R(\tau_i)
\end{equation}
where $c > 0$ is a positive constant. The resulting degree growth function is
\begin{equation}
\label{eq_degree_growth_r}
k_i(\tau_i, \eta_i) = \rme^{c \eta_i r(\tau_i)}
\end{equation}
where $r(\tau):=\int_0^{\tau} R(t)\,\der t$. By recognizing $c = 1/\theta$, we recover \eref{eq_degree_growth} as in the old formalism, hence the new formalism is consistent with the old one.
 
Now, for a given fitness distribution $\rho(\eta)$, we introduce function $h(\tau)$ as the \textit{average degree growth} of a node at age $\tau$,
\begin{equation}
h(\tau) = c \int\der\eta\,\rho(\eta) \times \eta k(\tau, \eta) R(\tau).
\end{equation}
We further introduce function $g(t)$ as the derivative of the network size $s$ with respect to the physical time $t$, $g(t) = \der s/\der t$. Hence $g(t)$ is the rate at which new nodes arrive in the system. Since each node is assumed to create one link, $g(t)$ is also the total degree increase of all existing nodes at time $t$. Considering the asymptotic behaviour ($t\to\infty$) of the network growth, we can now write $g$ as the convolution of $h$ and $g$ itself,
\begin{equation}
\label{eq_tiv}
g(t) = \int_{0}^{t}\der\tau\, h(\tau) g(t-\tau).
\end{equation}
Here we have the number of new links on the left side and the same quantity, expressed through degree increase of the existing nodes, on the right side.
Note so far we have not assumed any functional form of $g$.

\eref{eq_tiv} is the core of our new formalism. It describes a linear time-invariant (LTI) system $\mathcal H$ \cite{hespanha2018linear} whose impulse response function is $h$. Its input function happens to be the same as its output function,
\begin{equation}
\label{eq_eigen_H}
g = \mathcal H g.
\end{equation}
In other words, $g$ is the eigenfunction of the LTI operator $\mathcal H$ and thus it is of the exponential form $g(t) \sim \rme^{\sigma t}$ where $\sigma\in\mathbb{R}$ because $g$ is real. The eigenvalues of $\mathcal H$ can be given by the Laplace transform of the impulse response $h$,
\begin{equation}
\hat{h}(\sigma) = \mathcal{L} \{ h(\tau) \} = \int_{0}^{\infty}h(\tau)\rme^{-\sigma\tau}\,\der\tau.
\end{equation}
In Eq.~\ref{eq_eigen_H}, the corresponding eigenvalue of $g$ is exactly~$1$. We can thus get the exponential growth rate of the network, $\sigma$, by solving
\begin{equation}
\label{eq_sigma}
\hat{h}(\sigma) = 1.
\end{equation}
When $\sigma > 0$, we recover the exponential growth of network size $s(t) = \rme^{\sigma t} / \sigma$ which is analogous to $s(t)=\rme^{\alpha t}$ imposed by hand in the previous sections. When $\sigma = 0$, we have $g(t)=1$ which implies the linear network growth $s(t)=t$ as in~\cite{medo2011temporal}. When $\sigma < 0$, the model is still in principle valid but outside the scope of this study, since it means that as time progresses, fewer and fewer nodes join the network.

\subsection{Breaking of the time-invariance}
\label{sec_breaking_ti}
The time-invariance of the system as a whole is broken when $\hat{h}(\sigma)= 1$ does not have a solution.
To explain this, we start with the new formalism of the Bianconi-Barab\'{a}si model~\cite{bianconi2001competition} where $\Pi_i \sim \eta_i k_i$ and no aging is present.
The time-invariant degree growth function is thus a special case of \eref{eq_degree_growth_r} where $r(\tau) \equiv \tau$, i.e.,
\begin{equation}
k_i(\tau_i, \eta_i) = \rme^{c \eta_i \tau_i}.
\end{equation}
One can realize that the constant $c$ is merely a time scaler and is free of choice here, so for simplicity we let $c = 1$.
The impulse response can be written as
\begin{equation}
h(\tau) = \int\der\eta\, \rho(\eta) \eta\, \rme^{\eta \tau}.
\end{equation}
Solving
\begin{equation}
\label{eq_laplace}
\hat h(\sigma) = \mathcal L \{ h(\tau) \} = \int\der\eta\, \rho(\eta) \frac{\eta}{\sigma-\eta} = 1
\end{equation}
gives us the exponential growth rate of the network $\sigma$.
Since the degree growth rate of every node must not surpass the growth rate of the entire network, we have an additional constraint $\sigma \geq \eta_{\text{max}}$ where $\eta_{\text{max}}$ is the maximum fitness.

Since $\hat{h}(\sigma)$ is a decreasing function of $\sigma$, the maximum value of $\hat{h}(\sigma)$ is achieved at $\sigma=\eta_{\text{max}}$. However, for some fitness distributions (an example being $\rho(\eta) = (\lambda+1)(1-\eta)^\lambda$ where $\eta \in [0, 1]$ and $\lambda > 1$),
$\hat{h}(\eta_{\text{max}})$ is still smaller than $1$ which is required by \eref{eq_sigma}, and consequently, $\hat h(\sigma)= 1$ does not have a solution.
When such fitness distributions are taken, the node with the leading fitness will eventually attract almost all edges (a ``winner-takes-all'' effect). The network growth is thus asymptotically approached by the maximum degree growth, i.e., $g\sim k_{\text{max}}$ and, in the case of the Bianconi-Barab\'{a}si model, $g(t)\sim \rme^{\eta_{\text{max}} t}$. 
This situation has been intensively studied in~\cite{bianconi2001bose}, where the authors have approached the problem using the formalism used to study the Bose-Einstein condensation, and the critical parameter $\lambda_{\mathrm{BE}}=1$ when the condensation arises can be obtained.
In fact, by mapping fitness $\eta$ to energy $\epsilon$ at temperature $T$ with $\eta=\rme^{-\epsilon/T}$, one can realize that our \eref{eq_laplace} is equivalent to Eq.~(10) in \cite{bianconi2001bose}.

With our new formalism, we can also address other cases in which a similar gelation phenomenon arises, for instance the superlinear preferential attachment $\Pi_i\sim k_i^\gamma$ with $\gamma > 1$, where eventually a single node connects to nearly all other nodes~\cite{krapivsky2000connectivity}. This can be seen from the fact that the time-invariant degree growth function
\begin{equation}
\label{eq_degree_superl}
k(\tau) = \big[(1-\gamma)c \tau+1\big]^{1/(1-\gamma)},
\end{equation}
resulting from the differential equation $\der k/\der\tau = c k^\gamma$ with $\gamma > 1$ and $k(0) = 1$, displays a finite-time divergence at $\tau=[c(\gamma-1)]^{-1}$.
As a result, $\hat h(\sigma)=\int_{0}^{\infty}k(\tau)\rme^{-\sigma\tau}\der\tau$ does not converge for any real value $\sigma$, hence $\hat{h}(\sigma)= 1$ lacks a solution in $\mathbb R$.

Similar breaking of the time-invariance does not occur in the presence of aging where $\lim_{\tau\to\infty} R(\tau) = 0$. To prove this, we first examine the convergence of $\hat h(\sigma)$,
\begin{equation}
\hat{h}(\sigma) = \mathcal{L} \{ h(\tau) \} =
\mathcal{L}\Big\{ c \int\der\eta\,\rho(\eta) \times \eta k(\tau, \eta) R(\tau) \Big\}.
\end{equation}
Using the linearity of the Laplace transform $\mathcal{L}$, we can rewrite the equation above as
\begin{equation}
\hat{h}(\sigma) =
c \int\der\eta\,\rho(\eta) \times \eta \, \mathcal{L} \{ k(\tau, \eta) R(\tau) \}.
\end{equation}
Hence, $\hat h(\sigma)$ converges if $\mathcal{L} \{ k(\tau, \eta) R(\tau) \}$ converges for all $\eta$. 
This condition can further reduce to solely the convergence of $\mathcal{L} \{ k(\tau, \eta_{\max}) \}$ where $\eta_{\max}$ is the maximum fitness, since (1) $k(\tau, \eta_{\max}) \geq k(\tau, \eta)$ for all $\eta$, and
(2) the aging function $R(\tau)$ is decreasing. Recalling Eq.~(\ref{eq_degree_growth_r}), we have 
\begin{equation}
\label{eq_laplace_convergence}
\mathcal{L} \{ k(\tau, \eta_{\max}) \} =
\int_{0}^{\infty} \rme^{c \eta_{\max} r(\tau)} \,\rme^{-\sigma\tau}\,\der \tau.
\end{equation}
We thus examine the ratio
\begin{equation}
\label{eq_degree_ratio}
\frac{\rme^{c \eta_{\max} r(\tau+1) - \sigma(\tau+1)}}{\rme^{c \eta_{\max} r(\tau) - \sigma\tau}} 
    = \rme^{c \eta_{\text{max}} (r(\tau+1) - r(\tau))} \cdot \rme^{-\sigma}.
\end{equation}
When taking the limit $\tau \to \infty$, since $r'(\tau) = R(\tau)$ and $\lim_{\tau\to\infty} R(\tau) = 0$, we see that $\rme^{c \eta_{\text{max}} (r(\tau+1) - r(\tau))}$ approaches $1$. Therefore the examined ratio is less than $1$ when $\sigma > 0$, which guarantees the convergence of $\mathcal{L} \{ k(\tau, \eta_{\max}) \}$ and, consequently, of $\hat h(\sigma)$. Since $\hat h(\sigma)$ is a continuous monotonic function of $\sigma$ in the range $(0, \infty)$, there is always one solution of $\hat h(\sigma) = 1$.

\subsection{Degree distributions}
To conclude the analytical study of the model, we now derive its degree distribution which will be used in the following section to compare with numerical simulations.
Let $P(K \geq k, t)$ denote the probability that a node has degree at least $k$ at time $t$.
Since the fitness distribution does not change with time, $P(K \geq k, t)$ can be written as
\begin{equation}
  \label{eq_degree_P_t}
  P(K \geq k, t) = \frac{\sum_{\eta} n(K \geq k, t, \eta)}{s}
\end{equation}
where $n(K \geq k, t, \eta)$ represents the number of nodes with fitness $\eta$ that have degree at least $k$ at time $t$, and $s$ is the network size which we assume to have the exponential form $s=\exp(\alpha t)$.
Since for a given $\eta$, the relation between $k$ and $\tau$ is monotonous and independent of $t$ (recalling~\eref{eq_degree_growth_r}), we can write $\tau$ as a function of $k$,
\begin{equation}
\label{eq_degree_inverse}
\tau(k, \eta) = r^{-1}(\frac{\log k}{c \eta}).
\end{equation}
\eref{eq_degree_P_t} can be then rewritten using the ``mean-field'' approximation~\cite{barabasi1999mean,bollobas2001degree} as
\begin{equation}
  P(K \geq k, t) = \frac{\int\der\eta\,\rho(\eta)\, \rme^{\alpha [t - r^{-1}(\frac{\log k}{c \eta})]}}{\rme^{\alpha t}}
\end{equation}
when the network is large enough, in particular in the limit $t \to \infty$. The only time-dependent term $\rme^{\alpha t}$ cancels out and we obtain
\begin{equation}
\label{eq_degree_CCDF}
P(K \geq k) = \int\der\eta\,\rho(\eta) \, \rme^{-\alpha r^{-1}(\frac{\log k}{c\eta})}.
\end{equation}
A stationary degree distribution $P(k)$ thus exists,
\begin{equation}
\label{eq_degree_PMF}
P(k) \approx P(K \geq k) - P(K \geq k + 1).
\end{equation}

\begin{figure*}
\centering
\subfloat[\label{fig_simulation_ba}]
{\includegraphics[scale=0.5]{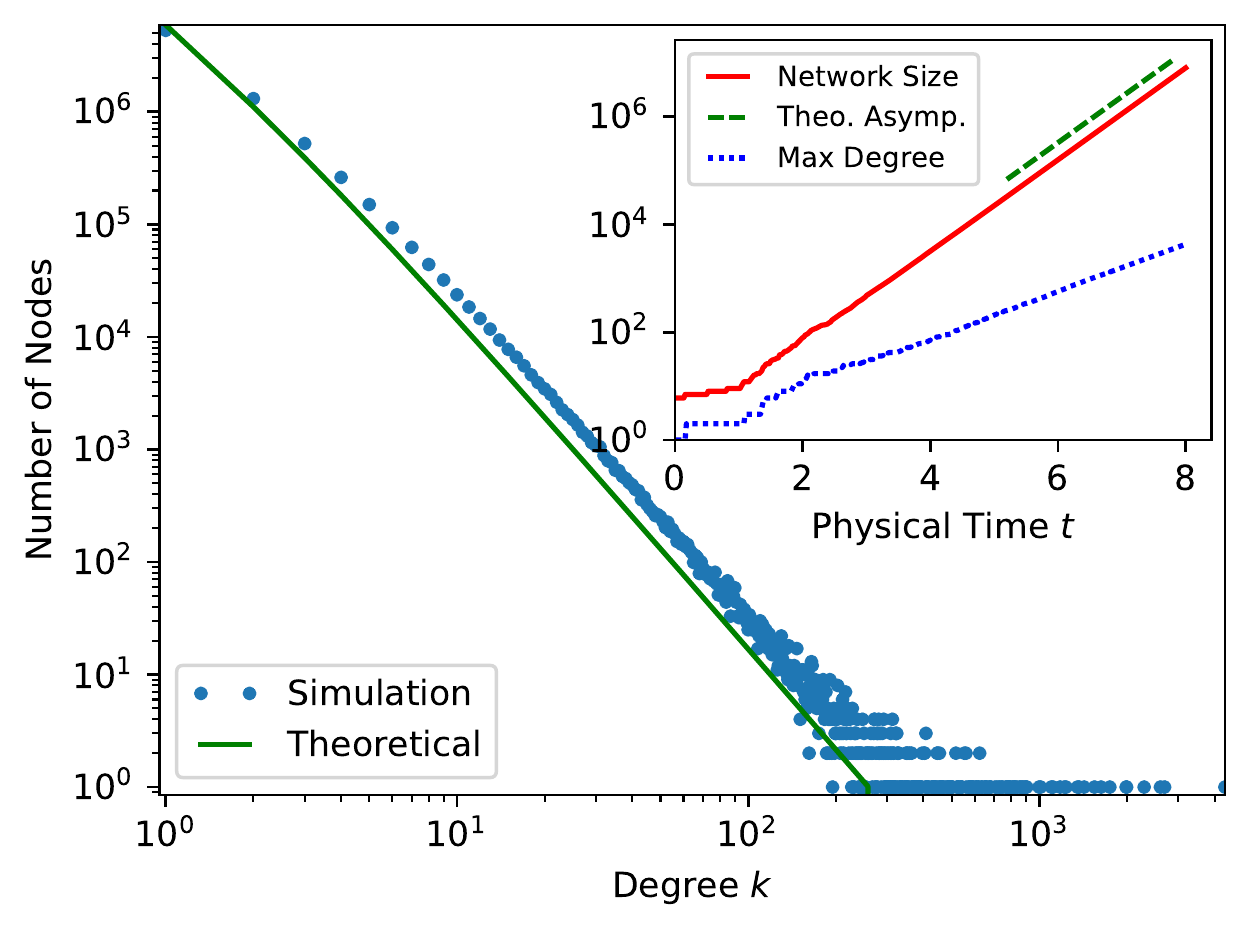}}
\quad
\subfloat[\label{fig_simulation_superl}]
{\includegraphics[scale=0.5]{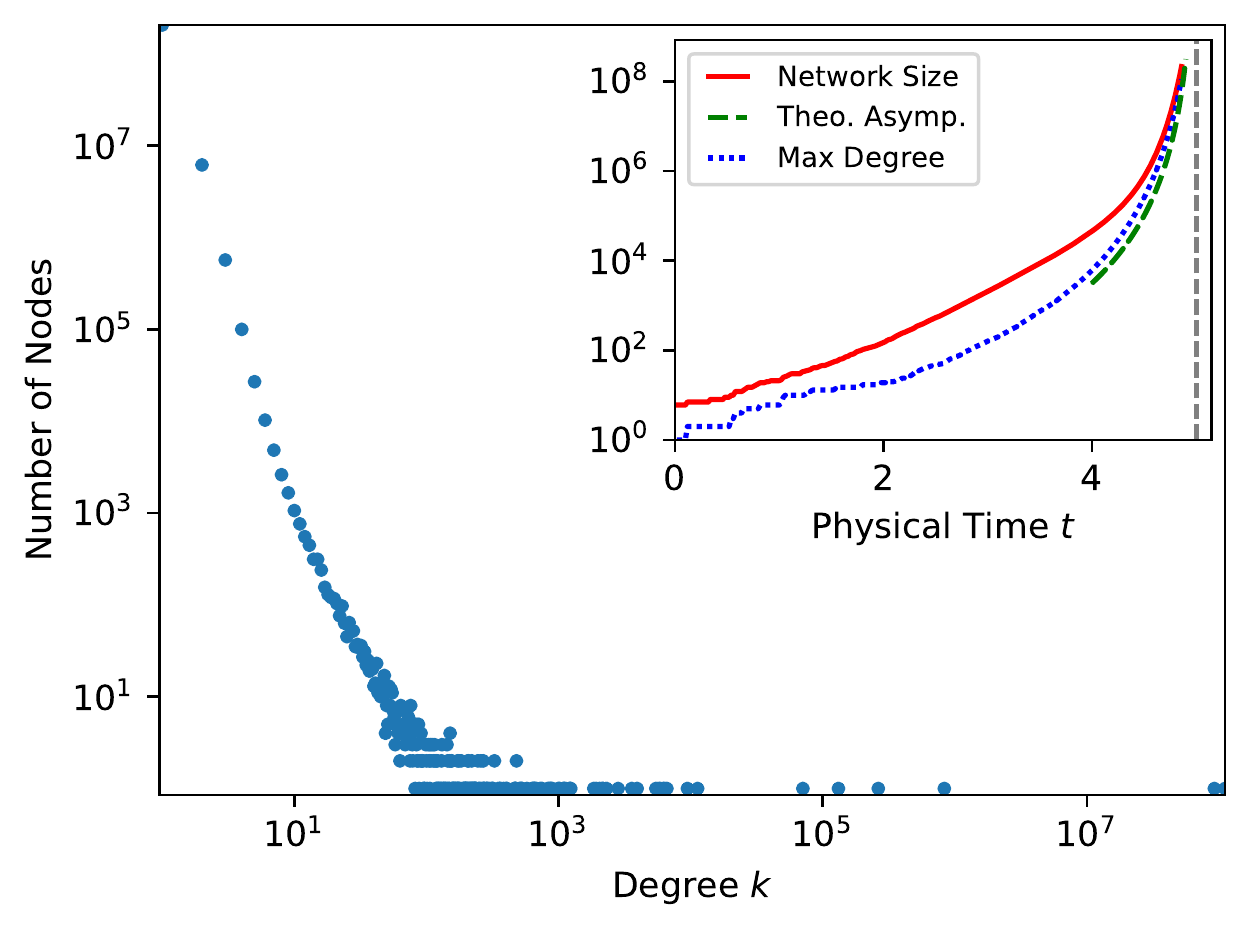}}\\
\subfloat[\label{fig_simulation_fitness_25}]
{\includegraphics[scale=0.5]{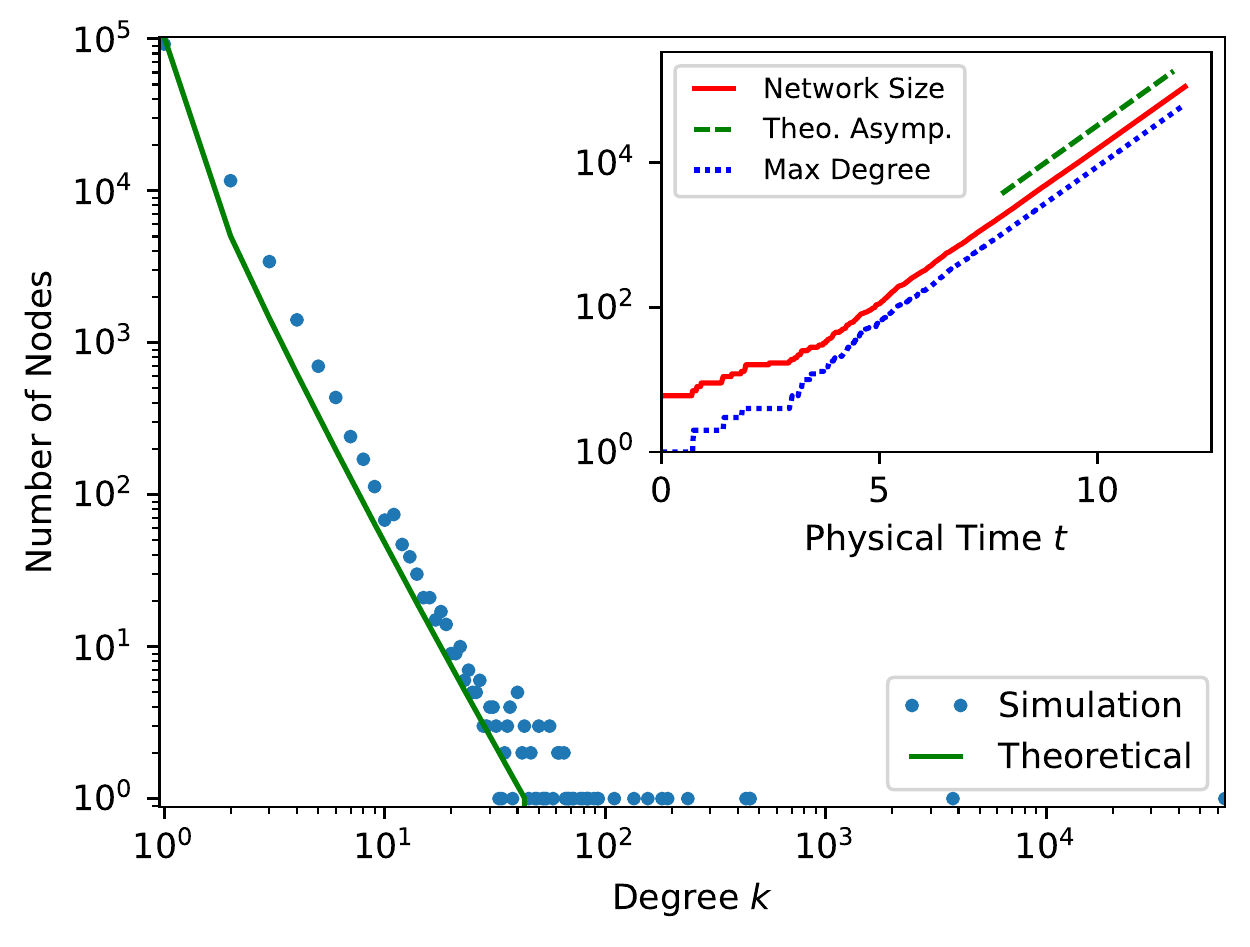}}
\quad
\subfloat[\label{fig_simulation_power_decay_09}]
{\includegraphics[scale=0.5]{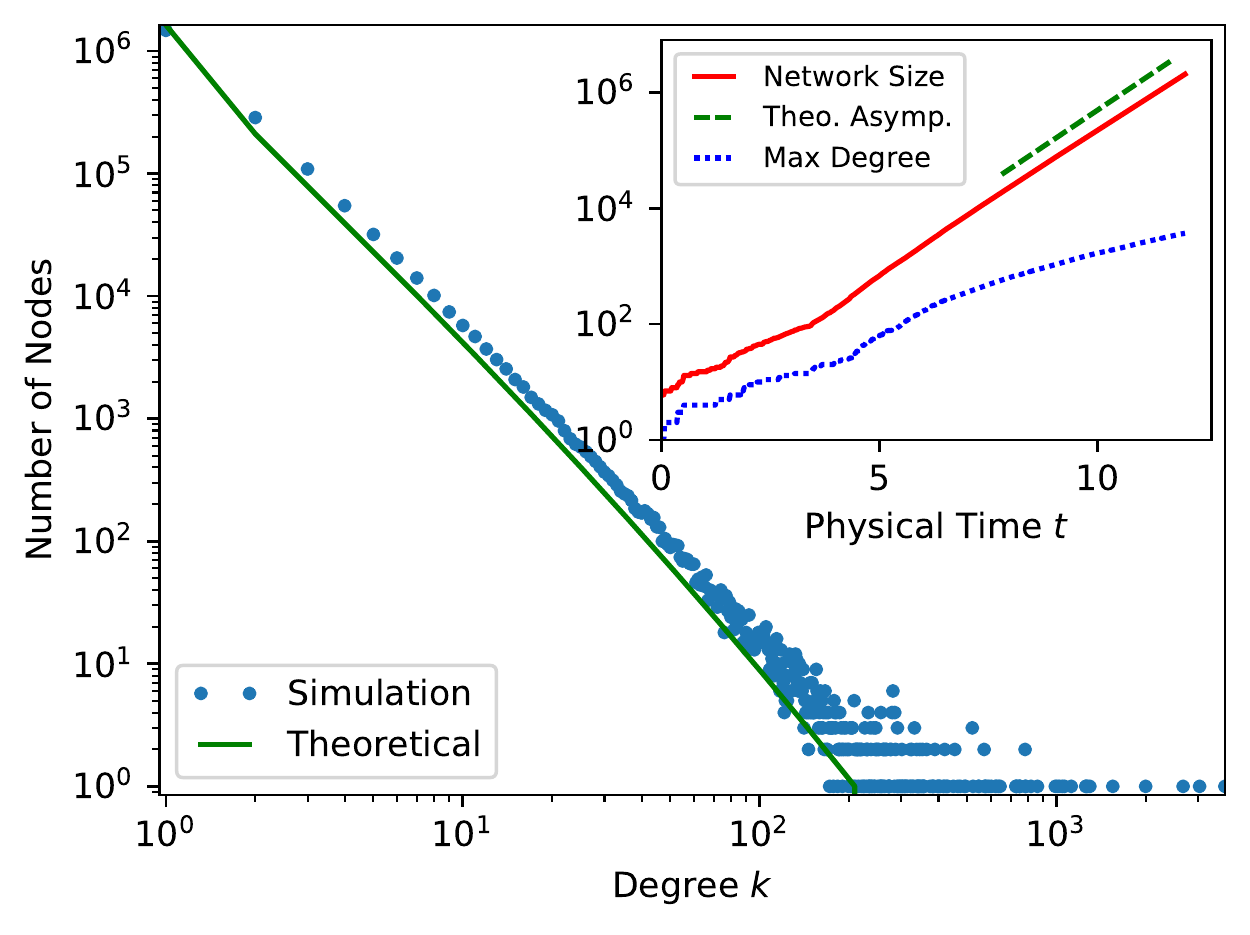}}
\caption{Simulation results for the degree distributions and the network growth, and their comparisons with analytical results:
(a) linear preferential attachment where $\der k / \der t \sim k$ as in the BA model;
(b) superlinear preferential attachment where $\der k / \der t \sim k^{1.2}$;
(c) Bianconi-Barab\'{a}si model where $\der k / \der t \sim k\eta$ and the fitness distribution is $\rho(\eta) \sim (1-\eta)^{2.5}$;
(d) preferential attachment with fitness and aging where $\der k / \der t \sim k\eta R(\tau)$, the fitness distribution is $\rho(\eta) \sim (1-\eta)^{2.5}$, and the aging function is $R(\tau) = 1/(\tau + 1)^{0.9}$.}
\label{fig_simulations}
\end{figure*}

\section{Simulations}
\label{sec_simulation}
To validate our analysis and the new formalism proposed in the last section, we grow synthetic networks with three major questions in mind:
\begin{enumerate}
\item Whether the time-invariant degree growth is consistent with the exponential network growth with the exponent $\sigma$ as predicted in \eref{eq_sigma},
\item Whether the winner-takes-all effect takes place when \eref{eq_sigma} lacks a solution,
\item Whether the model produces degree distributions given by \eref{eq_degree_CCDF}.
\end{enumerate}

In our simulations, we do not directly control the network size growth. Instead, the growth curve is left to be observed and compared with the model's analytical prediction.
Synthetic networks have initially six nodes with degree one each. Time runs in short time steps of size $\Delta t=0.02$ to limit the effects of time discretization. The degree increase of each node is drawn from the Poisson distribution with the mean increase given by \eref{eq_degree_pde}. A new node with degree one is added to the network whenever the degree of an existing node is increased by one. In this way, we effectively enforce the time-invariance of the degree growth and have the possibility to observe the emergent network growth.

\subsection{Results}
Simulation results shown in the insets of Fig.~\ref{fig_simulations} demonstrate that the emerging network growth in all cases eventually matches the theoretical prediction. When \eref{eq_sigma} has a solution $\sigma$, the network size exhibits an exponential growth with the exponent $\sigma$ (Fig.~\ref{fig_simulation_ba}, \ref{fig_simulation_power_decay_09}). In the case of the Bose-Einstein condensation, the network size grows with the exponent $\eta_{\max}$ (Fig.~\ref{fig_simulation_fitness_25}). Fig.~\ref{fig_simulation_superl} shows the superlinear preferential attachment which results in a network growth that is not exponentially bounded and can be approximated by the theoretical maximum degree growth curve $k = (5/(5-t))^5$ which follows from \eref{eq_degree_superl} for $\gamma=1.2$. As a result, the network size approaches infinity when $t=5$ (indicated with the vertical dashed line).

An important signature of the winner-takes-all effect is that the maximum degree eventually dominates the network growth, taking a fixed fraction of the network size.
In our new formalism, this happens when \eref{eq_sigma} lacks a solution, in the case of the superlinear preferential attachment (Fig.~\ref{fig_simulation_superl}) as well as in the Bianconi-Barab\'{a}si model when the Bose-Einstein condensation occurs~\cite{bianconi2001bose}, \ie, $\lambda \geq \lambda_\mathrm{BE} = 1$ (Fig.~\ref{fig_simulation_fitness_25}). As we have proven, in the presence of a diminishing aging function, there can be no winner-takes-all effect although $\lambda > \lambda_\mathrm{BE}$ (Fig.~\ref{fig_simulation_power_decay_09}).

For the degree distributions, the slopes of our theoretical results based on \eref{eq_degree_CCDF} match the synthetic results. Depending on the parameters, some networks have power-law shaped, well-defined long tail degree distributions (Fig.~\ref{fig_simulation_power_decay_09}). For the superlinear preferential attachment (Fig.~\ref{fig_simulation_superl}), since the network growth is not exponential, \eref{eq_degree_CCDF} does not apply and no stationary degree distribution is shown. This is in line with the known conclusion that superlinear preferential attachment networks lack an asymptotic stationary degree distribution~\cite{krapivsky2000connectivity}. In the case of the Bose-Einstein condensation (Fig.~\ref{fig_simulation_fitness_25}), there are some ``winners'' with large degree values, yet the slope of the estimated degree distribution still matches the simulation result for low degrees.

\section{Conclusion}
\label{sec:conclusion}
We provide here a comprehensive analysis of the effect of the accelerating network growth on the resulting networks created by various preferential attachment models. Such accelerated growth, albeit common in real systems, is typically neglected when analyzing network growth models. We find that instead of being an unnecessary nuisance, the form of the network growth is an important component which together with preferential attachment, fitness, and aging shapes the network.

Building on the observation that the average node degree growth in two different citation networks is time-invariant, we formulate a formalism which allows us to take the degree growth time-invariance as the first principle and study the emerging network properties.
The time-invariance of the degree growth is a natural property in networks that eventually reach ``stationary'' growth: their old and new nodes are alike in the way how their degree grows and saturates.
We use the new formalism to show that only two forms of network growth are compatible with the time-invariant degree growth: a uniform growth that is assumed by most network models and an exponential growth that is often found in real data. The simultaneous presence of time-invariant degree growth and an exponential network growth can be thus seen as empirical confirmation of these two patterns being self-consistent in growing networks with preferential attachment.
The new formalism naturally connects various network growth settings that have been previously studied separately: the Bose-Einstein condensation in a model with preferential attachment and fitness, a similar condensation in for superlinear preferential attachment, and the absence of such a condensation in the presence of aging.

Several questions remain open for future research. The exponential growth of the network size cannot sustain forever due to the limited number of potential nodes~\cite{suh2009singularity}, so it has to eventually slow down. Such a slowdown can be realized by relaxing the model assumptions by, for instance, changing the fitness distribution with time whilst still maintaining the time-invariant degree growth. Another possibility is to relax the time-invariance of the degree growth by allowing the parameter $c$ in \eref{eq_degree_pde} to vary. The exact form of the resulting network growth and its relation to the degree distribution are also interesting to study.

We have based our observations on citation networks. The studied model thus limits itself to no edge removal and edge creation only at node arrival. Besides, its preferential attachment process only considers local information of nodes, such as degree, fitness and aging. Lifting one or more of these limitations would enable extensions of the model to work with more general networks.

\begin{figure*}[t!]
\centering
\includegraphics[scale=0.5]{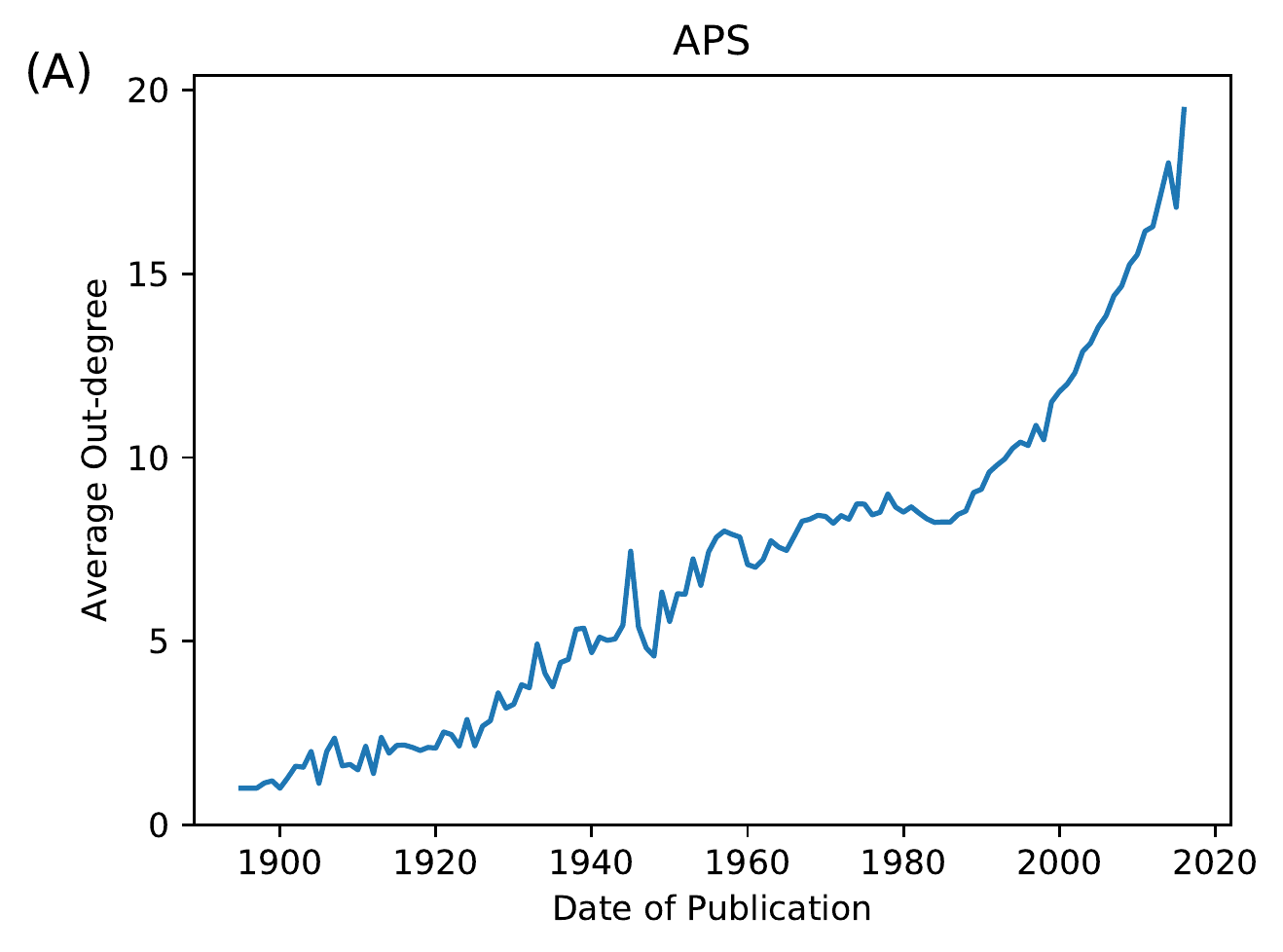}
\quad
\includegraphics[scale=0.5]{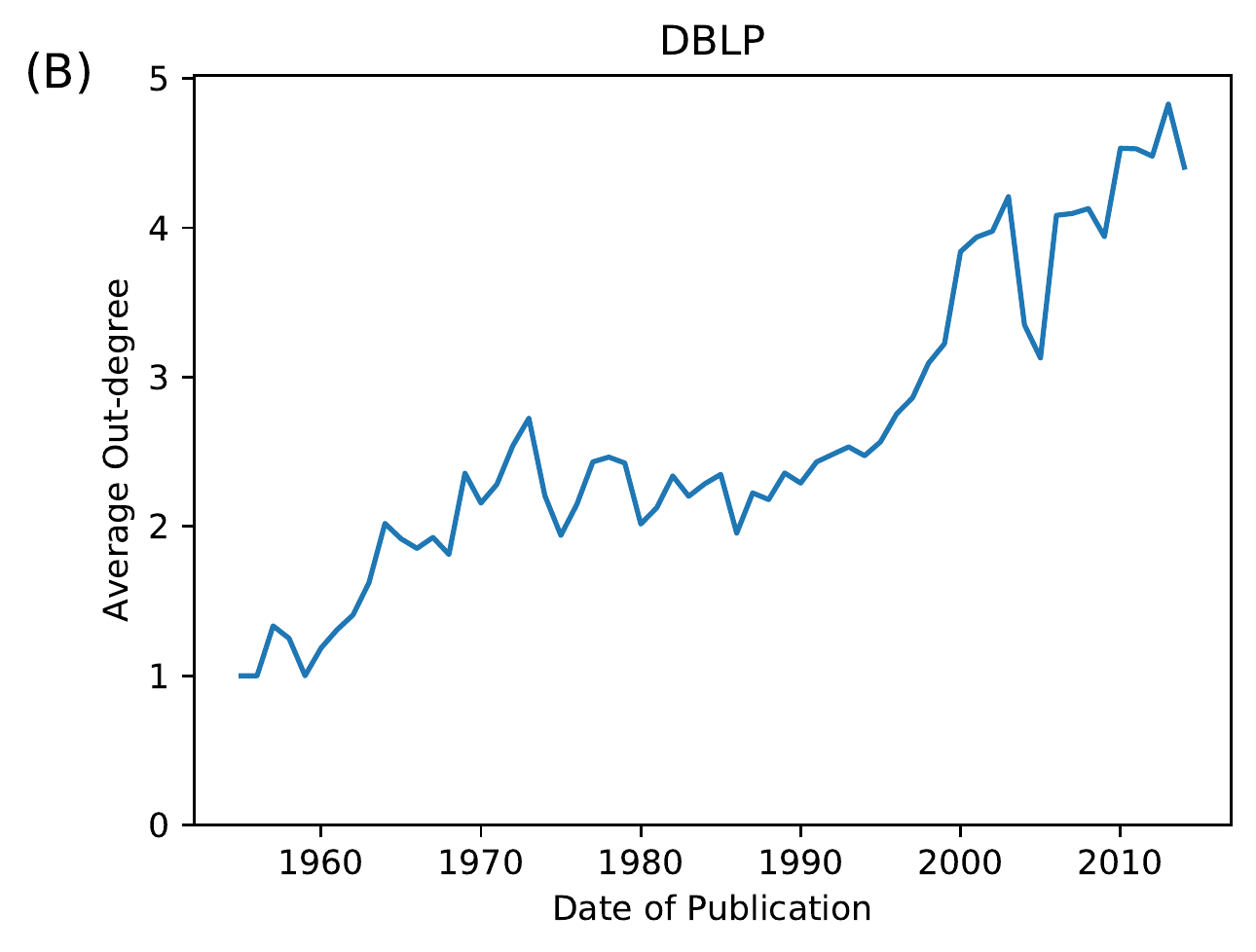}
\caption{The average out-degree of papers published in different years in (A) the APS data and (B) the DBLP data.}
\label{fig_outdegr}
\end{figure*}

\begin{figure*}[t!]
\centering
\includegraphics[scale=0.5]{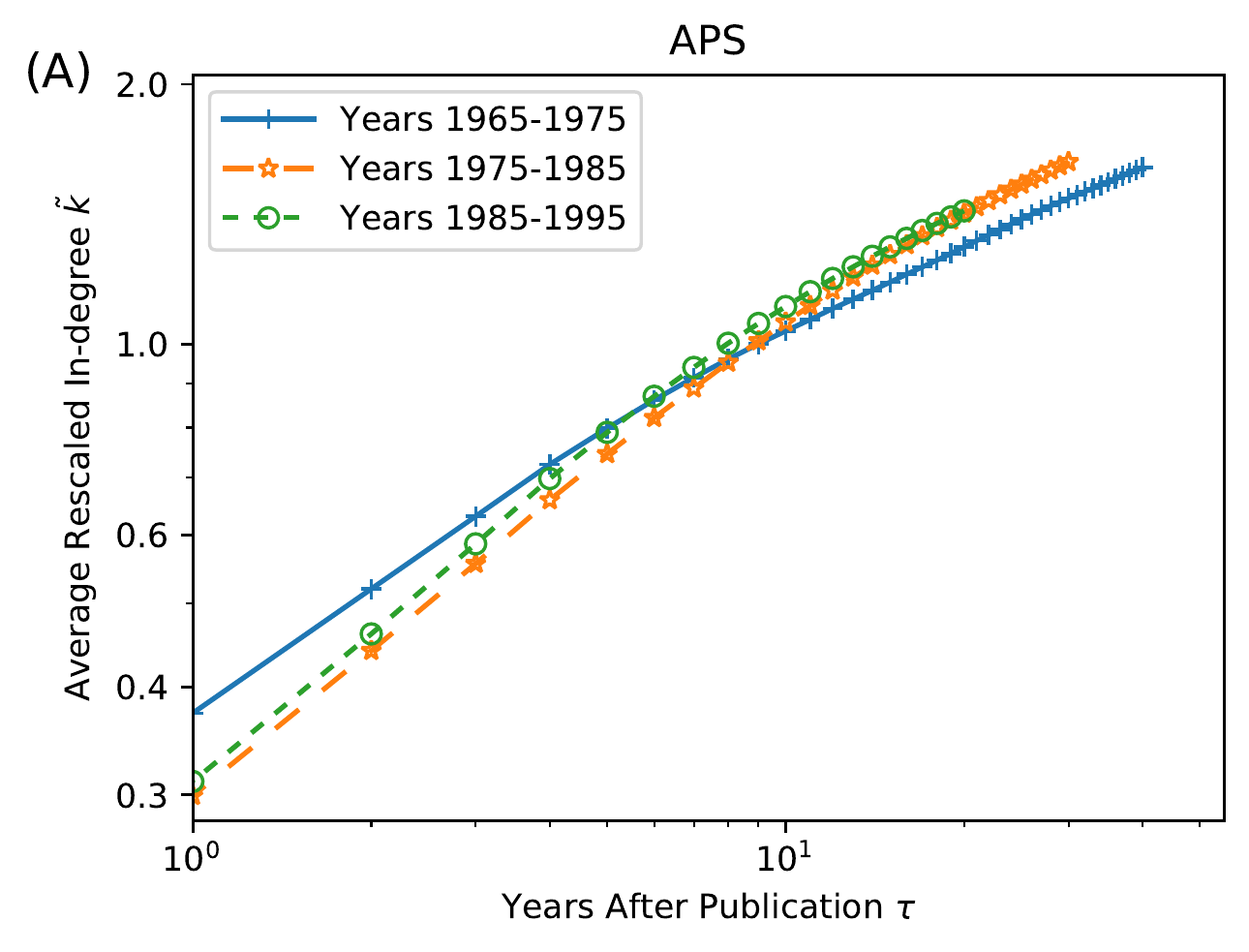}
\quad
\includegraphics[scale=0.5]{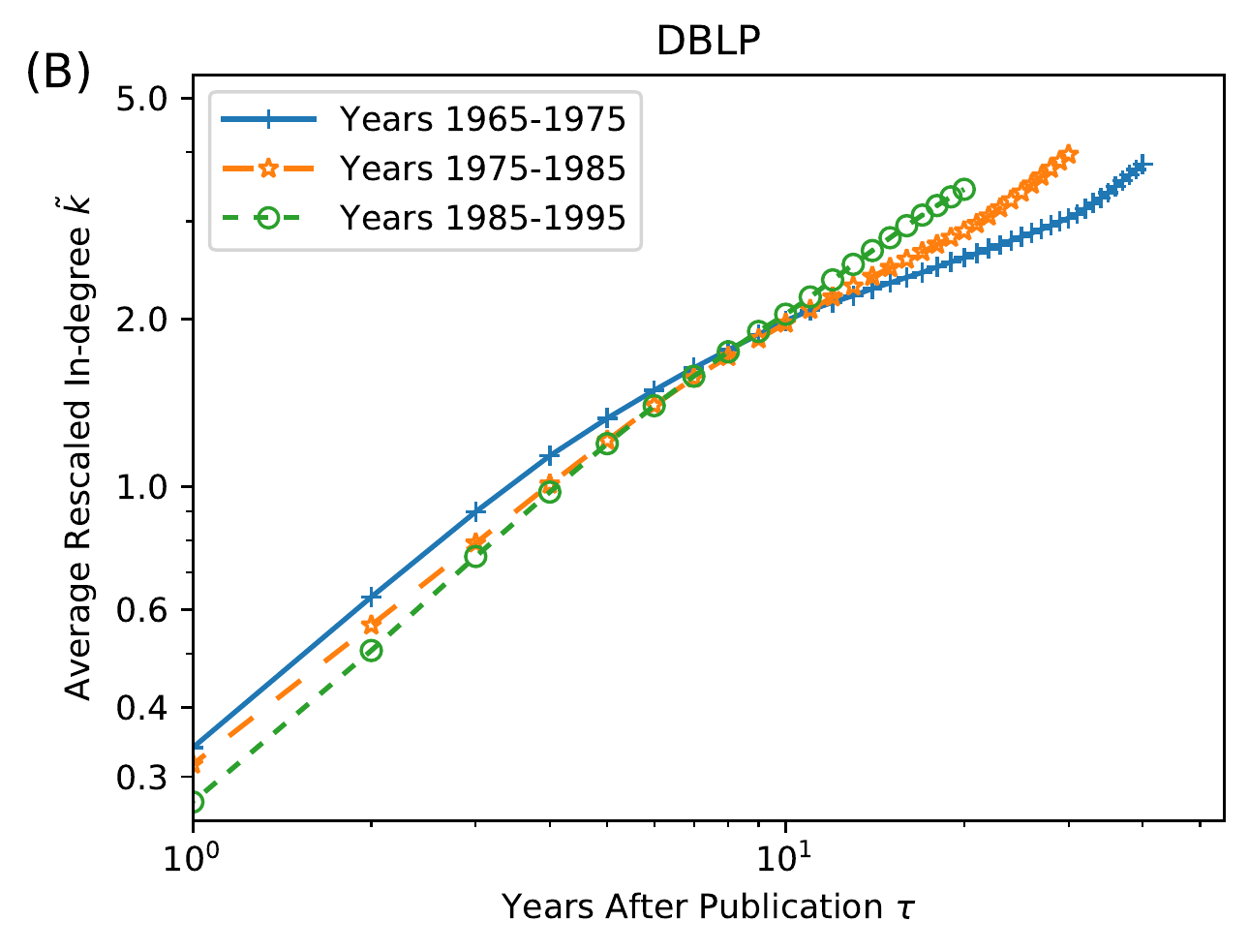}
\caption{The average number of citations as a function of the paper age, \textit{rescaled by the average out-degree in the years when the citations have been received}. Papers are divided in three groups by their publication year.}
\label{fig_rescaled}
\end{figure*}

\begin{acknowledgments}
We would like to thank the APS for providing us the citation data. This work is supported by the EU Horizon 2020 project CUTLER (\url{www.cutler-h2020.eu}) under contract No.~770469, the Swiss National Science Foundation (Grant No.~200020-156188) and the National Natural Science Foundation of China (Grant No.~11850410444). We appreciate helpful discussions with Sarah de Nigris.
\end{acknowledgments}

\appendix*
\label{appendix}
\section{Varying average out-degree in empirical data}
The growth of the average out-degree of papers with time (see Fig.~\ref{fig_outdegr} for the results in the two studied datasets) can be included in the model but it would come at the cost of increasing the model complexity. Instead, we limited the impact of the varying average out-degree on the empirical observations presented in Fig.~\ref{fig_exp_growth} by focusing on the period 1965--1995 during which the average out-degree changes relatively little.
One possible way to further limit such impact is to measure the in-degree growth using rescaled in-degree which divides the number of new citations in year $y$ by the average out-degree in this year and sums the contributions from individual years.
Fig.~\ref{fig_rescaled} shows that rescaled in-degree yields similar time-invariant growth patterns as measured using simple in-degree. 
In particular, the average rescaled in-degree $\tilde k$ in 10 years after publication are 1.04, 1.06 and 1.11 (APS), and 1.99, 1.97 and 2.04 (DBLP), respectively, for the three time periods shown in the figure. In comparison with Fig.~\ref{fig_exp_growth}, the growth curves are power-law over a broader range of paper age $\tau$.

%\nocite{*}
\bibliography{refs}

\begin{thebibliography}{29}%
\makeatletter
\providecommand \@ifxundefined [1]{%
 \@ifx{#1\undefined}
}%
\providecommand \@ifnum [1]{%
 \ifnum #1\expandafter \@firstoftwo
 \else \expandafter \@secondoftwo
 \fi
}%
\providecommand \@ifx [1]{%
 \ifx #1\expandafter \@firstoftwo
 \else \expandafter \@secondoftwo
 \fi
}%
\providecommand \natexlab [1]{#1}%
\providecommand \enquote  [1]{``#1''}%
\providecommand \bibnamefont  [1]{#1}%
\providecommand \bibfnamefont [1]{#1}%
\providecommand \citenamefont [1]{#1}%
\providecommand \href@noop [0]{\@secondoftwo}%
\providecommand \href [0]{\begingroup \@sanitize@url \@href}%
\providecommand \@href[1]{\@@startlink{#1}\@@href}%
\providecommand \@@href[1]{\endgroup#1\@@endlink}%
\providecommand \@sanitize@url [0]{\catcode `\\12\catcode `\$12\catcode
  `\&12\catcode `\#12\catcode `\^12\catcode `\_12\catcode `\%12\relax}%
\providecommand \@@startlink[1]{}%
\providecommand \@@endlink[0]{}%
\providecommand \url  [0]{\begingroup\@sanitize@url \@url }%
\providecommand \@url [1]{\endgroup\@href {#1}{\urlprefix }}%
\providecommand \urlprefix  [0]{URL }%
\providecommand \Eprint [0]{\href }%
\providecommand \doibase [0]{https://doi.org/}%
\providecommand \selectlanguage [0]{\@gobble}%
\providecommand \bibinfo  [0]{\@secondoftwo}%
\providecommand \bibfield  [0]{\@secondoftwo}%
\providecommand \translation [1]{[#1]}%
\providecommand \BibitemOpen [0]{}%
\providecommand \bibitemStop [0]{}%
\providecommand \bibitemNoStop [0]{.\EOS\space}%
\providecommand \EOS [0]{\spacefactor3000\relax}%
\providecommand \BibitemShut  [1]{\csname bibitem#1\endcsname}%
\let\auto@bib@innerbib\@empty
%</preamble>
\bibitem [{\citenamefont {Barab{\'a}si}\ and\ \citenamefont
  {Albert}(1999)}]{barabasi1999emergence}%
  \BibitemOpen
  \bibfield  {author} {\bibinfo {author} {\bibfnamefont {A.-L.}\ \bibnamefont
  {Barab{\'a}si}}\ and\ \bibinfo {author} {\bibfnamefont {R.}~\bibnamefont
  {Albert}},\ }\bibfield  {title} {\bibinfo {title} {Emergence of scaling in
  random networks},\ }\href@noop {} {\bibfield  {journal} {\bibinfo  {journal}
  {Science}\ }\textbf {\bibinfo {volume} {286}},\ \bibinfo {pages} {509}
  (\bibinfo {year} {1999})}\BibitemShut {NoStop}%
\bibitem [{\citenamefont {Barab{\'a}si}\ \emph {et~al.}(2016)\citenamefont
  {Barab{\'a}si} \emph {et~al.}}]{barabasi2016network}%
  \BibitemOpen
  \bibfield  {author} {\bibinfo {author} {\bibfnamefont {A.-L.}\ \bibnamefont
  {Barab{\'a}si}} \emph {et~al.},\ }\href@noop {} {\emph {\bibinfo {title}
  {Network Science}}}\ (\bibinfo  {publisher} {Cambridge University Press},\
  \bibinfo {year} {2016})\BibitemShut {NoStop}%
\bibitem [{\citenamefont {Newman}(2018)}]{newman2018networks}%
  \BibitemOpen
  \bibfield  {author} {\bibinfo {author} {\bibfnamefont {M.}~\bibnamefont
  {Newman}},\ }\href@noop {} {\emph {\bibinfo {title} {Networks}}}\ (\bibinfo
  {publisher} {Oxford University Press},\ \bibinfo {year} {2018})\BibitemShut
  {NoStop}%
\bibitem [{\citenamefont {Adamic}\ and\ \citenamefont
  {Huberman}(2000)}]{adamic2000power}%
  \BibitemOpen
  \bibfield  {author} {\bibinfo {author} {\bibfnamefont {L.~A.}\ \bibnamefont
  {Adamic}}\ and\ \bibinfo {author} {\bibfnamefont {B.~A.}\ \bibnamefont
  {Huberman}},\ }\bibfield  {title} {\bibinfo {title} {Power-law distribution
  of the world wide web},\ }\href@noop {} {\bibfield  {journal} {\bibinfo
  {journal} {Science}\ }\textbf {\bibinfo {volume} {287}},\ \bibinfo {pages}
  {2115} (\bibinfo {year} {2000})}\BibitemShut {NoStop}%
\bibitem [{\citenamefont {Medo}\ \emph {et~al.}(2011)\citenamefont {Medo},
  \citenamefont {Cimini},\ and\ \citenamefont {Gualdi}}]{medo2011temporal}%
  \BibitemOpen
  \bibfield  {author} {\bibinfo {author} {\bibfnamefont {M.}~\bibnamefont
  {Medo}}, \bibinfo {author} {\bibfnamefont {G.}~\bibnamefont {Cimini}},\ and\
  \bibinfo {author} {\bibfnamefont {S.}~\bibnamefont {Gualdi}},\ }\bibfield
  {title} {\bibinfo {title} {Temporal effects in the growth of networks},\
  }\href@noop {} {\bibfield  {journal} {\bibinfo  {journal} {Physical Review
  Letters}\ }\textbf {\bibinfo {volume} {107}},\ \bibinfo {pages} {238701}
  (\bibinfo {year} {2011})}\BibitemShut {NoStop}%
\bibitem [{\citenamefont {Jeong}\ \emph {et~al.}(2003)\citenamefont {Jeong},
  \citenamefont {N{\'e}da},\ and\ \citenamefont
  {Barab{\'a}si}}]{jeong2003measuring}%
  \BibitemOpen
  \bibfield  {author} {\bibinfo {author} {\bibfnamefont {H.}~\bibnamefont
  {Jeong}}, \bibinfo {author} {\bibfnamefont {Z.}~\bibnamefont {N{\'e}da}},\
  and\ \bibinfo {author} {\bibfnamefont {A.-L.}\ \bibnamefont {Barab{\'a}si}},\
  }\bibfield  {title} {\bibinfo {title} {Measuring preferential attachment in
  evolving networks},\ }\href@noop {} {\bibfield  {journal} {\bibinfo
  {journal} {EPL (Europhysics Letters)}\ }\textbf {\bibinfo {volume} {61}},\
  \bibinfo {pages} {567} (\bibinfo {year} {2003})}\BibitemShut {NoStop}%
\bibitem [{\citenamefont {Newman}(2001)}]{newman2001clustering}%
  \BibitemOpen
  \bibfield  {author} {\bibinfo {author} {\bibfnamefont {M.~E.}\ \bibnamefont
  {Newman}},\ }\bibfield  {title} {\bibinfo {title} {Clustering and
  preferential attachment in growing networks},\ }\href@noop {} {\bibfield
  {journal} {\bibinfo  {journal} {Physical Review E}\ }\textbf {\bibinfo
  {volume} {64}},\ \bibinfo {pages} {025102} (\bibinfo {year}
  {2001})}\BibitemShut {NoStop}%
\bibitem [{\citenamefont {Capocci}\ \emph {et~al.}(2006)\citenamefont
  {Capocci}, \citenamefont {Servedio}, \citenamefont {Colaiori}, \citenamefont
  {Buriol}, \citenamefont {Donato}, \citenamefont {Leonardi},\ and\
  \citenamefont {Caldarelli}}]{capocci2006preferential}%
  \BibitemOpen
  \bibfield  {author} {\bibinfo {author} {\bibfnamefont {A.}~\bibnamefont
  {Capocci}}, \bibinfo {author} {\bibfnamefont {V.~D.}\ \bibnamefont
  {Servedio}}, \bibinfo {author} {\bibfnamefont {F.}~\bibnamefont {Colaiori}},
  \bibinfo {author} {\bibfnamefont {L.~S.}\ \bibnamefont {Buriol}}, \bibinfo
  {author} {\bibfnamefont {D.}~\bibnamefont {Donato}}, \bibinfo {author}
  {\bibfnamefont {S.}~\bibnamefont {Leonardi}},\ and\ \bibinfo {author}
  {\bibfnamefont {G.}~\bibnamefont {Caldarelli}},\ }\bibfield  {title}
  {\bibinfo {title} {Preferential attachment in the growth of social networks:
  The internet encyclopedia {W}ikipedia},\ }\href@noop {} {\bibfield  {journal}
  {\bibinfo  {journal} {Physical Review E}\ }\textbf {\bibinfo {volume} {74}},\
  \bibinfo {pages} {036116} (\bibinfo {year} {2006})}\BibitemShut {NoStop}%
\bibitem [{\citenamefont {Bianconi}\ and\ \citenamefont
  {Barab{\'a}si}(2001{\natexlab{a}})}]{bianconi2001competition}%
  \BibitemOpen
  \bibfield  {author} {\bibinfo {author} {\bibfnamefont {G.}~\bibnamefont
  {Bianconi}}\ and\ \bibinfo {author} {\bibfnamefont {A.-L.}\ \bibnamefont
  {Barab{\'a}si}},\ }\bibfield  {title} {\bibinfo {title} {Competition and
  multiscaling in evolving networks},\ }\href@noop {} {\bibfield  {journal}
  {\bibinfo  {journal} {EPL (Europhysics Letters)}\ }\textbf {\bibinfo {volume}
  {54}},\ \bibinfo {pages} {436} (\bibinfo {year}
  {2001}{\natexlab{a}})}\BibitemShut {NoStop}%
\bibitem [{\citenamefont {Dorogovtsev}\ \emph {et~al.}(2000)\citenamefont
  {Dorogovtsev}, \citenamefont {Mendes},\ and\ \citenamefont
  {Samukhin}}]{dorogovtsev2000structure}%
  \BibitemOpen
  \bibfield  {author} {\bibinfo {author} {\bibfnamefont {S.~N.}\ \bibnamefont
  {Dorogovtsev}}, \bibinfo {author} {\bibfnamefont {J.~F.~F.}\ \bibnamefont
  {Mendes}},\ and\ \bibinfo {author} {\bibfnamefont {A.~N.}\ \bibnamefont
  {Samukhin}},\ }\bibfield  {title} {\bibinfo {title} {Structure of growing
  networks with preferential linking},\ }\href@noop {} {\bibfield  {journal}
  {\bibinfo  {journal} {Physical Review Letters}\ }\textbf {\bibinfo {volume}
  {85}},\ \bibinfo {pages} {4633} (\bibinfo {year} {2000})}\BibitemShut
  {NoStop}%
\bibitem [{\citenamefont {Albert}\ and\ \citenamefont
  {Barab{\'a}si}(2002)}]{albert2002statistical}%
  \BibitemOpen
  \bibfield  {author} {\bibinfo {author} {\bibfnamefont {R.}~\bibnamefont
  {Albert}}\ and\ \bibinfo {author} {\bibfnamefont {A.-L.}\ \bibnamefont
  {Barab{\'a}si}},\ }\bibfield  {title} {\bibinfo {title} {Statistical
  mechanics of complex networks},\ }\href@noop {} {\bibfield  {journal}
  {\bibinfo  {journal} {Reviews of Modern Physics}\ }\textbf {\bibinfo {volume}
  {74}},\ \bibinfo {pages} {47} (\bibinfo {year} {2002})}\BibitemShut {NoStop}%
\bibitem [{\citenamefont {Clauset}\ \emph {et~al.}(2009)\citenamefont
  {Clauset}, \citenamefont {Shalizi},\ and\ \citenamefont
  {Newman}}]{clauset2009power}%
  \BibitemOpen
  \bibfield  {author} {\bibinfo {author} {\bibfnamefont {A.}~\bibnamefont
  {Clauset}}, \bibinfo {author} {\bibfnamefont {C.~R.}\ \bibnamefont
  {Shalizi}},\ and\ \bibinfo {author} {\bibfnamefont {M.~E.}\ \bibnamefont
  {Newman}},\ }\bibfield  {title} {\bibinfo {title} {Power-law distributions in
  empirical data},\ }\href@noop {} {\bibfield  {journal} {\bibinfo  {journal}
  {SIAM Review}\ }\textbf {\bibinfo {volume} {51}},\ \bibinfo {pages} {661}
  (\bibinfo {year} {2009})}\BibitemShut {NoStop}%
\bibitem [{\citenamefont {Newman}(2009)}]{newman2009first}%
  \BibitemOpen
  \bibfield  {author} {\bibinfo {author} {\bibfnamefont {M.~E.}\ \bibnamefont
  {Newman}},\ }\bibfield  {title} {\bibinfo {title} {The first-mover advantage
  in scientific publication},\ }\href@noop {} {\bibfield  {journal} {\bibinfo
  {journal} {EPL (Europhysics Letters)}\ }\textbf {\bibinfo {volume} {86}},\
  \bibinfo {pages} {68001} (\bibinfo {year} {2009})}\BibitemShut {NoStop}%
\bibitem [{\citenamefont {Dorogovtsev}\ and\ \citenamefont
  {Mendes}(2001)}]{dorogovtsev2001effect}%
  \BibitemOpen
  \bibfield  {author} {\bibinfo {author} {\bibfnamefont {S.~N.}\ \bibnamefont
  {Dorogovtsev}}\ and\ \bibinfo {author} {\bibfnamefont {J.~F.~F.}\
  \bibnamefont {Mendes}},\ }\bibfield  {title} {\bibinfo {title} {Effect of the
  accelerating growth of communications networks on their structure},\
  }\href@noop {} {\bibfield  {journal} {\bibinfo  {journal} {Physical Review
  E}\ }\textbf {\bibinfo {volume} {63}},\ \bibinfo {pages} {025101} (\bibinfo
  {year} {2001})}\BibitemShut {NoStop}%
\bibitem [{\citenamefont {Parolo}\ \emph {et~al.}(2015)\citenamefont {Parolo},
  \citenamefont {Pan}, \citenamefont {Ghosh}, \citenamefont {Huberman},
  \citenamefont {Kaski},\ and\ \citenamefont
  {Fortunato}}]{parolo2015attention}%
  \BibitemOpen
  \bibfield  {author} {\bibinfo {author} {\bibfnamefont {P.~D.~B.}\
  \bibnamefont {Parolo}}, \bibinfo {author} {\bibfnamefont {R.~K.}\
  \bibnamefont {Pan}}, \bibinfo {author} {\bibfnamefont {R.}~\bibnamefont
  {Ghosh}}, \bibinfo {author} {\bibfnamefont {B.~A.}\ \bibnamefont {Huberman}},
  \bibinfo {author} {\bibfnamefont {K.}~\bibnamefont {Kaski}},\ and\ \bibinfo
  {author} {\bibfnamefont {S.}~\bibnamefont {Fortunato}},\ }\bibfield  {title}
  {\bibinfo {title} {Attention decay in science},\ }\href@noop {} {\bibfield
  {journal} {\bibinfo  {journal} {Journal of Informetrics}\ }\textbf {\bibinfo
  {volume} {9}},\ \bibinfo {pages} {734} (\bibinfo {year} {2015})}\BibitemShut
  {NoStop}%
\bibitem [{\citenamefont {Gagen}\ and\ \citenamefont
  {Mattick}(2005)}]{gagen2005accelerating}%
  \BibitemOpen
  \bibfield  {author} {\bibinfo {author} {\bibfnamefont {M.~J.}\ \bibnamefont
  {Gagen}}\ and\ \bibinfo {author} {\bibfnamefont {J.}~\bibnamefont
  {Mattick}},\ }\bibfield  {title} {\bibinfo {title} {Accelerating,
  hyperaccelerating, and decelerating networks},\ }\href@noop {} {\bibfield
  {journal} {\bibinfo  {journal} {Physical Review E}\ }\textbf {\bibinfo
  {volume} {72}},\ \bibinfo {pages} {016123} (\bibinfo {year}
  {2005})}\BibitemShut {NoStop}%
\bibitem [{\citenamefont {Bianconi}\ and\ \citenamefont
  {Barab{\'a}si}(2001{\natexlab{b}})}]{bianconi2001bose}%
  \BibitemOpen
  \bibfield  {author} {\bibinfo {author} {\bibfnamefont {G.}~\bibnamefont
  {Bianconi}}\ and\ \bibinfo {author} {\bibfnamefont {A.-L.}\ \bibnamefont
  {Barab{\'a}si}},\ }\bibfield  {title} {\bibinfo {title} {Bose-einstein
  condensation in complex networks},\ }\href@noop {} {\bibfield  {journal}
  {\bibinfo  {journal} {Physical Review Letters}\ }\textbf {\bibinfo {volume}
  {86}},\ \bibinfo {pages} {5632} (\bibinfo {year}
  {2001}{\natexlab{b}})}\BibitemShut {NoStop}%
\bibitem [{\citenamefont {Krapivsky}\ \emph {et~al.}(2000)\citenamefont
  {Krapivsky}, \citenamefont {Redner},\ and\ \citenamefont
  {Leyvraz}}]{krapivsky2000connectivity}%
  \BibitemOpen
  \bibfield  {author} {\bibinfo {author} {\bibfnamefont {P.~L.}\ \bibnamefont
  {Krapivsky}}, \bibinfo {author} {\bibfnamefont {S.}~\bibnamefont {Redner}},\
  and\ \bibinfo {author} {\bibfnamefont {F.}~\bibnamefont {Leyvraz}},\
  }\bibfield  {title} {\bibinfo {title} {Connectivity of growing random
  networks},\ }\href@noop {} {\bibfield  {journal} {\bibinfo  {journal}
  {Physical Review Letters}\ }\textbf {\bibinfo {volume} {85}},\ \bibinfo
  {pages} {4629} (\bibinfo {year} {2000})}\BibitemShut {NoStop}%
\bibitem [{\citenamefont {Tang}\ \emph {et~al.}(2008)\citenamefont {Tang},
  \citenamefont {Zhang}, \citenamefont {Yao}, \citenamefont {Li}, \citenamefont
  {Zhang},\ and\ \citenamefont {Su}}]{aminer}%
  \BibitemOpen
  \bibfield  {author} {\bibinfo {author} {\bibfnamefont {J.}~\bibnamefont
  {Tang}}, \bibinfo {author} {\bibfnamefont {J.}~\bibnamefont {Zhang}},
  \bibinfo {author} {\bibfnamefont {L.}~\bibnamefont {Yao}}, \bibinfo {author}
  {\bibfnamefont {J.}~\bibnamefont {Li}}, \bibinfo {author} {\bibfnamefont
  {L.}~\bibnamefont {Zhang}},\ and\ \bibinfo {author} {\bibfnamefont
  {Z.}~\bibnamefont {Su}},\ }\bibfield  {title} {\bibinfo {title}
  {Arnet{M}iner: Extraction and mining of academic social networks},\ }in\
  \href@noop {} {\emph {\bibinfo {booktitle} {Proc. KDD'08}}}\ (\bibinfo
  {publisher} {ACM},\ \bibinfo {year} {2008})\ pp.\ \bibinfo {pages}
  {990--998}\BibitemShut {NoStop}%
\bibitem [{\citenamefont {Ley}(2002)}]{ley2002dblp}%
  \BibitemOpen
  \bibfield  {author} {\bibinfo {author} {\bibfnamefont {M.}~\bibnamefont
  {Ley}},\ }\bibfield  {title} {\bibinfo {title} {The {DBLP} computer science
  bibliography: Evolution, research issues, perspectives},\ }in\ \href@noop {}
  {\emph {\bibinfo {booktitle} {String Processing and Information Retrieval}}}\
  (\bibinfo {organization} {Springer},\ \bibinfo {year} {2002})\ pp.\ \bibinfo
  {pages} {1--10}\BibitemShut {NoStop}%
\bibitem [{\citenamefont {Caldarelli}\ \emph {et~al.}(2002)\citenamefont
  {Caldarelli}, \citenamefont {Capocci}, \citenamefont {{De Los Rios}},\ and\
  \citenamefont {Munoz}}]{caldarelli2002scale}%
  \BibitemOpen
  \bibfield  {author} {\bibinfo {author} {\bibfnamefont {G.}~\bibnamefont
  {Caldarelli}}, \bibinfo {author} {\bibfnamefont {A.}~\bibnamefont {Capocci}},
  \bibinfo {author} {\bibfnamefont {P.}~\bibnamefont {{De Los Rios}}},\ and\
  \bibinfo {author} {\bibfnamefont {M.~A.}\ \bibnamefont {Munoz}},\ }\bibfield
  {title} {\bibinfo {title} {Scale-free networks from varying vertex intrinsic
  fitness},\ }\href@noop {} {\bibfield  {journal} {\bibinfo  {journal}
  {Physical Review Letters}\ }\textbf {\bibinfo {volume} {89}},\ \bibinfo
  {pages} {258702} (\bibinfo {year} {2002})}\BibitemShut {NoStop}%
\bibitem [{\citenamefont {Medo}(2014)}]{medo2014statistical}%
  \BibitemOpen
  \bibfield  {author} {\bibinfo {author} {\bibfnamefont {M.}~\bibnamefont
  {Medo}},\ }\bibfield  {title} {\bibinfo {title} {Statistical validation of
  high-dimensional models of growing networks},\ }\href@noop {} {\bibfield
  {journal} {\bibinfo  {journal} {Physical Review E}\ }\textbf {\bibinfo
  {volume} {89}},\ \bibinfo {pages} {032801} (\bibinfo {year}
  {2014})}\BibitemShut {NoStop}%
\bibitem [{\citenamefont {Wang}\ \emph {et~al.}(2013)\citenamefont {Wang},
  \citenamefont {Song},\ and\ \citenamefont {Barab{\'a}si}}]{wang127}%
  \BibitemOpen
  \bibfield  {author} {\bibinfo {author} {\bibfnamefont {D.}~\bibnamefont
  {Wang}}, \bibinfo {author} {\bibfnamefont {C.}~\bibnamefont {Song}},\ and\
  \bibinfo {author} {\bibfnamefont {A.-L.}\ \bibnamefont {Barab{\'a}si}},\
  }\bibfield  {title} {\bibinfo {title} {Quantifying long-term scientific
  impact},\ }\href@noop {} {\bibfield  {journal} {\bibinfo  {journal}
  {Science}\ }\textbf {\bibinfo {volume} {342}},\ \bibinfo {pages} {127}
  (\bibinfo {year} {2013})}\BibitemShut {NoStop}%
\bibitem [{\citenamefont {Kleinberg}\ \emph {et~al.}(1999)\citenamefont
  {Kleinberg}, \citenamefont {Kumar}, \citenamefont {Raghavan}, \citenamefont
  {Rajagopalan},\ and\ \citenamefont {Tomkins}}]{kleinberg1999web}%
  \BibitemOpen
  \bibfield  {author} {\bibinfo {author} {\bibfnamefont {J.~M.}\ \bibnamefont
  {Kleinberg}}, \bibinfo {author} {\bibfnamefont {R.}~\bibnamefont {Kumar}},
  \bibinfo {author} {\bibfnamefont {P.}~\bibnamefont {Raghavan}}, \bibinfo
  {author} {\bibfnamefont {S.}~\bibnamefont {Rajagopalan}},\ and\ \bibinfo
  {author} {\bibfnamefont {A.~S.}\ \bibnamefont {Tomkins}},\ }\bibfield
  {title} {\bibinfo {title} {The web as a graph: measurements, models, and
  methods},\ }in\ \href@noop {} {\emph {\bibinfo {booktitle} {International
  Computing and Combinatorics Conference}}}\ (\bibinfo {organization}
  {Springer},\ \bibinfo {year} {1999})\ pp.\ \bibinfo {pages}
  {1--17}\BibitemShut {NoStop}%
\bibitem [{\citenamefont {Sun}\ \emph {et~al.}(2018)\citenamefont {Sun},
  \citenamefont {Staab},\ and\ \citenamefont {Karimi}}]{sun2018decay}%
  \BibitemOpen
  \bibfield  {author} {\bibinfo {author} {\bibfnamefont {J.}~\bibnamefont
  {Sun}}, \bibinfo {author} {\bibfnamefont {S.}~\bibnamefont {Staab}},\ and\
  \bibinfo {author} {\bibfnamefont {F.}~\bibnamefont {Karimi}},\ }\bibfield
  {title} {\bibinfo {title} {Decay of relevance in exponentially growing
  networks},\ }in\ \href@noop {} {\emph {\bibinfo {booktitle} {Proc.
  WebSci'18}}}\ (\bibinfo  {publisher} {ACM},\ \bibinfo {year} {2018})\ pp.\
  \bibinfo {pages} {343--351}\BibitemShut {NoStop}%
\bibitem [{\citenamefont {Hespanha}(2018)}]{hespanha2018linear}%
  \BibitemOpen
  \bibfield  {author} {\bibinfo {author} {\bibfnamefont {J.~P.}\ \bibnamefont
  {Hespanha}},\ }\href@noop {} {\emph {\bibinfo {title} {Linear Systems
  Theory}}}\ (\bibinfo  {publisher} {Princeton university press},\ \bibinfo
  {year} {2018})\BibitemShut {NoStop}%
\bibitem [{\citenamefont {Barab{\'a}si}\ \emph {et~al.}(1999)\citenamefont
  {Barab{\'a}si}, \citenamefont {Albert},\ and\ \citenamefont
  {Jeong}}]{barabasi1999mean}%
  \BibitemOpen
  \bibfield  {author} {\bibinfo {author} {\bibfnamefont {A.-L.}\ \bibnamefont
  {Barab{\'a}si}}, \bibinfo {author} {\bibfnamefont {R.}~\bibnamefont
  {Albert}},\ and\ \bibinfo {author} {\bibfnamefont {H.}~\bibnamefont
  {Jeong}},\ }\bibfield  {title} {\bibinfo {title} {Mean-field theory for
  scale-free random networks},\ }\href@noop {} {\bibfield  {journal} {\bibinfo
  {journal} {Physica A: Statistical Mechanics and its Applications}\ }\textbf
  {\bibinfo {volume} {272}},\ \bibinfo {pages} {173} (\bibinfo {year}
  {1999})}\BibitemShut {NoStop}%
\bibitem [{\citenamefont {Bollob{\'a}s}\ \emph {et~al.}(2001)\citenamefont
  {Bollob{\'a}s}, \citenamefont {Riordan}, \citenamefont {Spencer},
  \citenamefont {Tusn{\'a}dy} \emph {et~al.}}]{bollobas2001degree}%
  \BibitemOpen
  \bibfield  {author} {\bibinfo {author} {\bibfnamefont {B.}~\bibnamefont
  {Bollob{\'a}s}}, \bibinfo {author} {\bibfnamefont {O.}~\bibnamefont
  {Riordan}}, \bibinfo {author} {\bibfnamefont {J.}~\bibnamefont {Spencer}},
  \bibinfo {author} {\bibfnamefont {G.}~\bibnamefont {Tusn{\'a}dy}}, \emph
  {et~al.},\ }\bibfield  {title} {\bibinfo {title} {The degree sequence of a
  scale-free random graph process},\ }\href@noop {} {\bibfield  {journal}
  {\bibinfo  {journal} {Random Structures \& Algorithms}\ }\textbf {\bibinfo
  {volume} {18}},\ \bibinfo {pages} {279} (\bibinfo {year} {2001})}\BibitemShut
  {NoStop}%
\bibitem [{\citenamefont {Suh}\ \emph {et~al.}(2009)\citenamefont {Suh},
  \citenamefont {Convertino}, \citenamefont {Chi},\ and\ \citenamefont
  {Pirolli}}]{suh2009singularity}%
  \BibitemOpen
  \bibfield  {author} {\bibinfo {author} {\bibfnamefont {B.}~\bibnamefont
  {Suh}}, \bibinfo {author} {\bibfnamefont {G.}~\bibnamefont {Convertino}},
  \bibinfo {author} {\bibfnamefont {E.~H.}\ \bibnamefont {Chi}},\ and\ \bibinfo
  {author} {\bibfnamefont {P.}~\bibnamefont {Pirolli}},\ }\bibfield  {title}
  {\bibinfo {title} {The singularity is not near: Slowing growth of
  {W}ikipedia},\ }in\ \href@noop {} {\emph {\bibinfo {booktitle} {Proc.
  WikiSym'09}}}\ (\bibinfo {organization} {ACM},\ \bibinfo {year} {2009})\
  p.~\bibinfo {pages} {8}\BibitemShut {NoStop}%
\end{thebibliography}%

\end{document}